\documentclass[aps,twocolumn,showpacs,amsmath,amssymb,superscriptaddress]{revtex4}

\usepackage{epsfig}
\usepackage{graphicx}
\usepackage{graphics}
\usepackage{dcolumn}
\usepackage{bm}
\usepackage{hyperref}

\begin{document}

\title{Improved initial data for black hole binaries by 
asymptotic matching of post-Newtonian and perturbed black hole solutions}

\author{Nicol\'as  Yunes}
\affiliation{Institute for Gravitational Physics and Geometry,
             Center for Gravitational Wave Physics,
             Department of Physics, The Pennsylvania State University,
             University Park, PA 16802-6300}

\author{Wolfgang Tichy}
\affiliation{Department of Physics, Florida Atlantic University,
             Boca Raton, FL  33431}

\date{$$Id: paper.tex,v 1.118 2006/07/11 11:19:29 wolf Exp $$}



%
\newcommand\be{\begin{equation}}
\newcommand\ba{\begin{eqnarray}}

\newcommand\ee{\end{equation}}
\newcommand\ea{\end{eqnarray}}
\newcommand\p{{\partial}}
\newcommand\remove{{{\bf{THIS FIG. OR EQS. COULD BE REMOVED}}}}
%

\begin{abstract}
  
We construct approximate initial data for non-spinning black hole binary
systems by asymptotically matching the 4-metrics of two tidally perturbed
Schwarzschild solutions in isotropic coordinates to a resummed
post-Newtonian 4-metric in ADMTT coordinates. The specific matching
procedure used here closely follows the calculation in~\cite{Yunes:2005nn},
and is performed in the so called buffer zone where both the post-Newtonian
and the perturbed Schwarzschild approximations hold. The result is that both
metrics agree in the buffer zone, up to the errors in the approximations.
However, since isotropic coordinates are very similar to ADMTT coordinates,
matching yields better results than in the previous
calculation~\cite{Yunes:2005nn}, where harmonic coordinates were used for
the post-Newtonian 4-metric.
In particular, not only does matching improve in the
buffer zone, but due to the similarity between ADMTT and isotropic
coordinates the two metrics are also close to each other near the black hole
horizons.
%
%
With the help of a transition function we also obtain a global smooth
4-metric which has errors on the order of the error introduced by the more
accurate of the two approximations we match. 
This global smoothed out 4-metric is obtained in ADMTT coordinates which are
not horizon penetrating. In addition, we construct a further coordinate
transformation that takes the 4-metric from global ADMTT coordinates to new
coordinates which are similar to Kerr-Schild coordinates near each black
hole, but which remain ADMTT further away from the black holes. These new
coordinates are horizon penetrating and lead, for example, to a lapse which
is everywhere positive on the $t=0$ slice. Such coordinates may be more
useful in numerical simulations.

\end{abstract}

\maketitle

\section{Introduction}
The modeling of binary black hole systems is essential for the
detection of gravitational waves both by space and earth-based
interferometers~\cite{Schutz99,LIGO_web,GEO_web,VIRGO_web,TAMA_web}.
Often this detection relies on matched filtering, which
consists of comparing accurate waveform templates to the
signal. In the strong field regime, these templates can be modeled
accurately enough only by numerical 
simulations~\cite{Miller:2005qu},
where one solves the Einstein equations subject to initial conditions
that determine the physical character of the system. The astrophysical
accuracy of the template generated via a numerical evolution will
depend on two main factors: the numerical error introduced by the
evolution and gravitational wave extraction codes; and the
astrophysical accuracy of the initial data set. In this paper we study
one particular approach to solve the initial data problem.
This approach was developed in Ref.~\cite{Yunes:2005nn} (henceforth paper
$1$) and is based on matching several approximate solutions of the 
Einstein equations.

There have been numerous efforts to find astrophysically accurate
initial data for binary systems
\cite{Cook94,York:1998hy,Matzner98a,Baumgarte00a,
Marronetti00,Marronetti00a,Cook:2000vr,Grandclement:2001ed,Baker02a,
Tichy:2002ec,Pfeiffer:2002xz,Tichy03a,Tichy:2003qi,
Hannam:2003tv,Bonning:2003im,Holst:2004wt,Ansorg:2004ds,Yo:2004ng,
Cook:2004kt,Hannam:2005rp},
which usually consist of the $3$-metric and extrinsic curvature.
Most methods rely on performing decompositions of the data to
separate the so called physical degrees of freedom from those 
constrained by the
Einstein equations and those associated with diffeomorphisms. These
physical degrees of freedom carry the gravitational wave content of
the data, and thus should be chosen to represent the astrophysical
scenario that is being described. However, the
exact solution to the Einstein equations for the $2$-body scenario
remains elusive. Therefore, the gravitational wave content of the
data is not known exactly.

One commonly used cure to this problem is to give an ansatz for these physical
degrees of freedom, which usually does not satisfy the constraints. This
ansatz is then projected onto the constraint-satisfying hypersurface by
solving the constraint equations for the quantities introduced by the
decomposition. The projected data will now satisfy the constraints, but it
is unclear whether such data are astrophysically accurate. First, some of
the assumptions that are used in the construction of the initial ansatz are
known to be physically inaccurate. For example, the spatial metric
is often assumed to be conformally flat, which is known to be incorrect at
$2$ post-Newtonian (PN) order \cite{Blanchet:2002av}, and which also
does not contain realistic tidal deformations near the black holes.
Furthermore, after projecting onto the constraint hypersurface,
the data will be physically different from the initial ansatz
\cite{Aguirregabiria:2001vk,Pfeiffer:2002xz,Tichy:2002ec}. 
It is thus unclear exactly what physical scenario these projected data
represent and whether or not they will be realistic.

Recent efforts have concentrated on using physical approximations to
construct initial data sets. One such effort can be called the
quasi-equilibrium
approach~\cite{Grandclement:2001ed,Tichy:2003qi,Cook:2004kt}
where one assumes that the quantities describing the initial data evolve on
a time scale much longer than the orbital timescale, when appropriate
coordinates are used. There are also approaches built
entirely from second post-Newtonian 
approximations~\cite{Tichy:2002ec,Nissanke:2005kp}. Even though 
\cite{Tichy:2002ec} also develops a complete method to project onto
the constraint hypersurface, both 
\cite{Tichy:2002ec} and \cite{Nissanke:2005kp} are not guaranteed
to be realistic close to the black holes, since 
post-Newtonian theory is in principle not valid close to black holes.
Another recent effort, which will be the subject
of this work, involves several analytic approximations that are both
physically accurate and close to the constraint hypersurface far and
close to the holes (see paper $1$).
Such initial data coming from analytic approximations can
then be evolved without solving the constraints, provided that
constraint violations are everywhere smaller than numerical error. A
perhaps more appealing alternative is to project these data onto the
constraint hypersurface. If these data are close enough to the
hypersurface, then any sensible projection algorithm should produce
constraint-satisfying data that are close to the original approximate
solution.

Regardless of how such data are implemented, there are reasons to believe
that such an approach will produce astrophysically realistic initial data.
First, since the data are built from
physical approximations, there are no assumptions that are
unrealistic. The only limit to the accuracy is given by the order
to which the approximations are taken. Moreover, the analytic control
provided by these physical approximations allows for the tracking of
errors in the physics, thus providing a means to measure the distance
to the exact initial data set. 
Note however, that our work described below will only be accurate up to
leading order in both the post-Newtonian and the tidal perturbation
expansions used. This means that at the currently
achieved expansion order, our initial data are by no means guaranteed
to be superior to the other approaches discussed above. Nevertheless,
our approach can be systematically extended to higher order.

Data that use analytic approximations valid on the entire hypersurface
are difficult to find because such approximations are usually valid
only in certain regions of the hypersurface. One method to construct
such data is by asymptotically matching two different physical
approximations in a region where they overlap. Asymptotic matching,
which for general relativity was developed
in~\cite{Burke,Burke-Thorne,Death:1974o,Death:1975,Thorne:1984mz},
consists of comparing the asymptotic approximations of two adjacent
solution inside of an overlap $4$-volume. By comparing these
asymptotic approximations, matching returns a map between coordinates
and parameters local to different regions, which forces adjacent
solutions to be asymptotic to each other inside the overlap region
called the buffer zone.  Since these solutions are close to each other
in the buffer zone, it is possible to construct transition functions
that merge these solutions, thus generating a smooth global metric.

Alvi \cite{Alvi:1999cw} attempted to apply matching to binary systems,
but instead of matching he actually performed patching, because he set
the physical approximations equal to each other at a $2$-surface. As a
result, Jansen and Br\"ugmann \cite{Bernd} found that Alvi's
$4$-metric, and in particular his extrinsic curvature, was hard to 
smooth with transition functions in the buffer zone, since the
solutions were not sufficiently close to each other. Jansen and
Br\"ugmann thus concluded that these large discontinuities in Alvi's
data renders them impractical for numerical implementation. In paper $1$,
binary systems were studied once more, but this time true asymptotic
matching was implemented, thus successfully generating solutions that
approach each other in the buffer zone. It was possible to smooth
these solutions and thus to create useful initial data.
The approach we used in paper $1$ was to match a PN metric
in harmonic gauge~\cite{Blanchet:2002av},
valid far from both black holes but less than a gravitational
wavelength from the center of mass (near zone), to a
perturbed Schwarzschild metric in isotropic coordinates, valid close
to the holes (inner zone). 

In this paper, we study the method developed in paper $1$ in more
detail and we investigate and improve the resulting data.  This
improvement is mainly due to using different coordinates in the PN
approximation. Instead of employing harmonic gauge, we now use a
post-Newtonian metric in ADMTT
gauge~\cite{Schafer:1986rd,Jaranowski97,Jaranowski:1997ky}.  We will
see that this leads to better matching.  The reason for this
improvement is that the 4-metric in ADMTT coordinates can be brought
into a form that is very close to Schwarzschild in isotropic
coordinates near each black hole. Therefore, we obtain a much smoother
match between this near zone metric and the inner zone one, which is
also given in isotropic coordinates (see Fig.~\ref{gxx}).  The results
presented in Sec.~\ref{secglobal} indeed show that matching with the
ADMTT PN near zone metric works better than with the harmonic PN near
zone metric. In addition, the ADMTT PN near zone metric is very
similar to the perturbed Schwarzschild metric even near the black hole
horizon, where PN approximations in principle break down. This
similarity is due to the fact that, we use a resummed version of the
ADMTT metric, obtained by adding higher order post-Newtonian terms.

Another piece of evidence that suggests that matching
works better between ADMTT and isotropic coordinates comes from
looking at Hamiltonian and momentum constraint violations.
In Sec.~\ref{constraints} we compare the Hamiltonian and momentum constraint
violations for the data of this paper and of paper $1$. We find
that the use of the ADMTT PN near zone metric leads to smaller
constraint violations than in paper $1$.
With the help of a transition function we can also obtain a global smooth
4-metric which has errors on the order of the error introduced by the more
accurate of the two approximations we match. 

By computing the constraint violations, we discover that both the data
set of this paper and that presented in paper $1$ might not be easily
implemented in numerical simulations. In particular, we observe that
the momentum constraint diverges near the apparent horizon of
each black hole. This makes it difficult or impossible
to excise a region inside each black hole, which contains all
points where divergences occur. This divergence arises due to the
choice of coordinates in the inner zone.
Since we use the $t=const$ slices of isotropic coordinates
as spatial slices, the lapse goes through zero near the horizon.
This zero in the lapse in turn leads to a blowup of some
components of the extrinsic curvature and the momentum constraint. In order
to circumvent this problem, we construct a map from isotropic coordinates
to new horizon-penetrating coordinates. A similar idea was presented
in Ref.~\cite{Alvi:2003pn}, but here we extend those ideas and provide
explicit expressions for the transformed inner zone metric
in a ready-to-implement form. These new
coordinates are identical Kerr-Schild coordinates inside the horizon and
become isotropic in the buffer zone. When this transformation is applied
to either the data set presented in this paper or that presented in
paper $1$, the lapse of the new $t=const$ slices is positive through the
horizon, eliminating the divergence in the extrinsic curvature. This new
coordinate system will make excision easier, without changing the physical
content of the data.

The paper is organized as follows. Sec.~\ref{scenario} explains how
the spacetime is divided into zones and how matching will be
implemented.  Sec.~\ref{near} describes the near zone post-Newtonian
metric, while Sec.~\ref{inner} focuses on the inner zone metric.
Sec.~\ref{matching} performs the matching and provides a map between
coordinates and parameters local to the near and inner zones.
Sec.~\ref{secglobal} gives explicit formulas for the global metric,
introduces the smoothing functions and also decomposes this $4$-metric
into initial data for numerical relativity.  Sec.~\ref{constraints}
compares the constraint violations of the initial data presented in
this paper to those of paper $1$. In Sec.~\ref{efcoords} we present an
additional coordinate transformation which can be used to construct
horizon penetrating coordinates. Sec.~\ref{conclusion} concludes and
points to future work. Throughout we use geometrized units, where
$G=c=1$.

\section{Division of spacetime into zones and matching in GR}
\label{scenario}
\begin{figure}
\includegraphics[scale=0.33,clip=true]{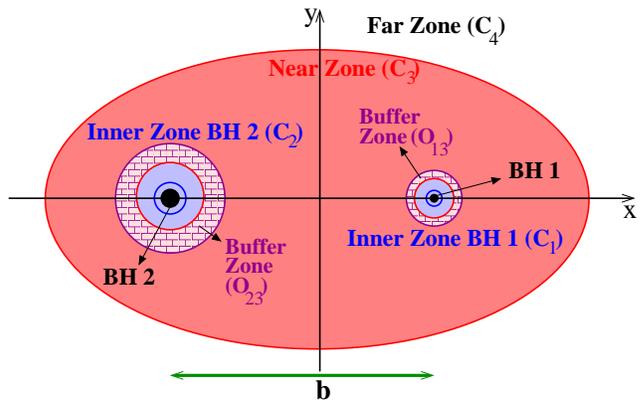}
\caption{\label{egg} Schematic diagram of the near zone (dark
  gray), inner zones (light gray) and buffer zones (checkered). The
  near zone extends up to the inner boundary of the buffer zone, while
  the inner zone extends from the outer boundary of the buffer zone up
  to the black hole. Therefore, the buffer zones are spherical shells where
  the near zone and inner zones overlap. BH~1 and BH~2 (black) are
  separated by distance $b$.}
\end{figure}

Consider a binary black hole spacetime and divide it into $4$ zones.
First, there are the so called inner zones ${\cal{C}}_1$ and
${\cal{C}}_2$ close to each black hole, where we can use black hole
perturbation theory to obtain an approximate solution to the Einstein
equations. These solutions are obtained under the assumption that the
black holes are separated far enough that the influence of black hole
$2$ is only a small perturbation near black hole $1$. Second, there is
the near zone ${\cal{C}}_3$ where PN theory should provide a good
approximation as long as the black holes do not move too fast. Finally
${\cal{C}}_4$ denotes the far zone where retardation effects matter.
These zones are shown in Fig.~\ref{egg} and we summarize them in Table
I.
\begin{table}[htb]
\begin{center}
\medskip
\begin{tabular}{||c|c||c|c||}
\hline
\hline
Zone & $r_{in}$ & $r_{out}$  & $\epsilon_{(n)}$ \\
\hline\hline
Inner zone BH $1$ (${\cal{C}}_1$) & 0 & $\ll b$ & $\bar{r}_1/b$ \\
Inner zone BH $2$ (${\cal{C}}_2$) & 0 & $\ll b$ & $\bar{r}_2/b$ \\
Near zone (${\cal{C}}_3$) & $\gg m_A$ & $\ll \lambda/2 \pi$ & $m_A/r_A$ \\
Far zone (${\cal{C}}_4$) & $\gg b $ & $\infty$ & $m_A/r_A$ \\
\hline\hline
\end{tabular}
\\[10pt]
{Table I: Description of the division of the spacetime into
  zones~\cite{Alvi:1999cw}.}
\end{center}

\end{table}
The quantities introduced in this table are defined as follows:
$r_{in}$ and $r_{out}$ are the approximate inner and outer boundary of
each zone respectively, as measured from the $A$th black hole ($A=1$
or $2$) in the coordinate system local to it; $\epsilon_{(n)}$ is the
expansion parameter used to find an approximate solution to the
Einstein equations in that zone; $m_A$ is the mass parameter of the
$A$th black hole in the near and far zones; $r_A$ and $\bar{r}_A$ are the
radial distances as measured from the $A$th hole in the near and inner
zones respectively; $\lambda$ is the typical gravitational wavelength;
and $b$ is the coordinate separation of the black holes. 

Inside of each zone we approximate the $4$-metric with an monovariate
expansion in $\epsilon_{(n)}$, which is defined in some coordinate
system, and which depends on parameters, like the mass of the black
holes or the orbital angular velocity. The inner zones are both
equipped with isotropic coordinates and have parameters ($M_A$,
$\Omega$), where the approximate solutions to the Einstein equations
are given by a perturbed Schwarzschild metric. On the other hand, the
near zone is equipped with ADMTT coordinates (ADMTT gauge), where the
approximate solution to the Einstein equations is a PN approximation
with parameters $m_A$ and $\omega$. This PN approximation is valid in
both ${\cal{C}}_3$ and ${\cal{C}}_4$ up to the order treated in this
paper.

As explained in paper $1$, asymptotic matching consists of comparing
the asymptotic expansions of adjacent approximate metrics in an
overlap region or buffer zone. These buffer zones are defined by the
intersection of the regions of validity of two adjacent metrics. The
$2$ buffer zones are delimited by the intersection of the inner zones
${\cal{C}}_1$ and ${\cal{C}}_2$ (of BHs $1$ and $2$), and the near
zone ${\cal{C}}_{3}$.  These buffer zones ${\cal{O}}_{12} =
{\cal{C}}_{1} \cap {\cal{C}}_{2}$ and ${\cal{O}}_{13} = {\cal{C}}_{1}
\cap {\cal{C}}_{3}$ can only be defined in the asymptotic
sense~\cite{Bender} via $m_A \ll r_A \ll b$ and are the 
regions in which asymptotic matching will be performed.  Formally,
there is a third buffer zone (${\cal{O}}_{34}$), given by the
intersection of the near and far zone, but since the ADMTT PN metric
at the order considered here is valid in both ${\cal{C}}_3$ and
${\cal{C}}_4$, it is unnecessary to perform any matching there.  The
asymptotic expansions of the approximate metrics in the buffer zones
${\cal{O}}_{A3}$ are bivariate, because they will depend on $2$
independent parameters: $m_A/r_A$ and $r_A/b$. Thus, the errors in the
asymptotic expansions valid in the buffer zones will be denoted as
$O(p,q)$, which stands for errors of $O(m/b)^{p}$ or errors of
$O(r_A/b)^q$.

Asymptotic matching then produces a coordinate and parameter mapping
between adjacent regions. In this paper, these maps will consist of a
transformation between isotropic and ADMTT coordinates, as well as a
set of conditions to relate $M_A$ and $\Omega$ to $m_A$ and $\omega$.
With these maps, we can construct a piece-wise global metric by
choosing a $3$-surface inside the buffer zone, where we join the
approximate solutions to the Einstein equations together. Small
discontinuities will be present in the global metric at the chosen
$3$-surface, due to its inherent piecewise nature, but they are of the
same order as the errors in the physical approximations, and thus
controllable by the size of the perturbation parameters. Furthermore,
if such parameters are chosen sufficiently small, these errors could,
in particular, become smaller than numerical discretization errors.
Ultimately, a $C^{\infty}$ metric is sought after, so, since these
discontinuities are small, the global metric can be smoothed by introducing
transition functions.

\section{Near Zone: An ADMTT post-Newtonian metric}
\label{near}

In this section, we present the near zone metric in ADMTT gauge valid
in ${\cal C}_3$ and ${\cal{C}}_4$ and expand it in the buffer zone
$\mathcal{O}_{13}$.  Analyzing matching in $\mathcal{O}_{13}$ will
suffice due to the symmetry of the problem. The coordinate-parameter
map for the overlap region $\mathcal{O}_{23}$ will later be given by a
simple symmetry transformation.

The ADMTT PN metric is obtained by solving the equations of motion for
a binary black hole system with Hamiltonian dynamics
\cite{Schafer:1986rd}. The Hamiltonian is expanded in a slow-motion,
weak-field approximation (PN) in ADMTT coordinates. This approach is
similar to the standard Lagrangian PN expansion implemented in paper
$1$ and there exists a mapping between the two methods
\cite{Blanchet:2002av}. However, they differ in that the Hamiltonian
formulation introduces a $3+1$ decomposition of the $4$-metric from
the start, and thus it provides PN expressions for the lapse, shift,
$3$-metric and the conjugate momentum.

We follow the conventions of Ref.~\cite{Schafer:1986rd}
where the black hole trajectories from the center of mass of the
system is given by
\be
\label{trajectories}
\vec{\xi}_1(t'') = \frac{m_2}{m} \vec{b}(t''), \qquad 
\vec{\xi}_2(t'') = -\frac{m_1}{m} \vec{b}(t'').
\ee
In Eq.~(\ref{trajectories}), $m_A$ and $m=m_1 + m_2$ stand for the
mass of BH A and the combined mass of the system in the ADMTT gauge,
whereas the separation vector $\vec{b}(t)$ is given by
\be
\label{b-def}
\vec{b}(t'') = b\left(\cos{\omega t''}, \sin{\omega t''}, 0\right).
\ee
Note that Eq.~(\ref{b-def}) assumes the black holes are initially in a
circular orbit, which is a sensible approximation in the late stages
of inspiral, since gravitational radiation will have circularized the
orbit. The angular velocity of this orbit is given by Eq.~(60) in
Ref.~\cite{Tichy:2002ec}:
\be 
\label{ang-vel-def}
\omega=\sqrt{m/b^3}\left[1+ \frac{1}{2} 
       \left(\frac{\mu}{m}-3\right) \frac{m}{b}\right],
\ee
with errors of $O(m/b)^{5/2}$. In Eq.~(\ref{ang-vel-def}), $\mu = m_1
m_2/m$ is the reduced mass of the system and $b$ is the norm of the
separation vector.  This angular velocity is valid in the ADMTT gauge,
which is different from the PN velocity used in paper $1$, since
the latter is valid only in harmonic coordinates.

We further define the radial vector pointing from either black hole to
a field point by 
\be
\vec{r}_A{}'' = (x''-\xi_A^x, y''-\xi_A^y, z''-\xi_A^z), 
\ee
where the double primed variables ($t'', x'', y'', z''$) are inertial
ADMTT coordinates measured from the center of mass.  With these
definitions, it is clear that $\vec{r}_1{}'' =\vec{r}_2{}'' +
\vec{b}$, where $\vec{b}$ points from black hole $1$ to $2$.
We also introduce the unit vector
\be
\vec{n}_A{}'' = \vec{r}_A{}''/r_A{}'' .
\ee
and the radial vector pointing from the center of mass to a field
point $r'' = (x'', y'', z'')$.

The particle's velocity vector is given by
\be
\vec{v}_A{}'' = \partial_t \vec{r}_A{}'' =
\left(-1\right)^{A+1}\frac{\mu}{m_A}  b \omega 
\left(-\sin{\omega  t''},\cos{\omega t''}, 0\right).
\ee

The post-Newtonian near zone metric in inertial ADMTT coordinates
(Eqs.~(5.~4)-(5.~6) in Ref.~\cite{Schafer:1986rd}) can then be written
as
\ba
\label{admtt-metric}
g_{i''j''}^{(3)} &=& \Psi^4 \delta_{ij} ,
\nonumber \\
g_{0''i''}^{(3)} &=& g_{i''j''}^{(3)} \beta^{(3) j''},
\nonumber \\
g_{0''0''}^{(3)} &=& g_{0''i''}^{(3)} \beta^{(3) i''} - (\alpha^{(3)})^2,
\nonumber \\
\ea
where we introduced a post-Newtonian conformal factor
\be
\label{Psi_ADMTT}
\Psi = 1 + \frac{m_1}{2 r_1''} + \frac{m_2}{2 r_2''}.
\ee
and where the post-Newtonian 
lapse and shift (Eqs.(5.4)-(5.6) in Ref.~\cite{Schafer:1986rd})
are written as
\be
\label{alpha_ADMTT}
\alpha^{(3)}= \frac{2-\Psi}{\Psi}
\ee
and
\ba
\label{beta_ADMTT}
\beta^{(3) i''} &=& 
\frac{m_1}{r_1''} \left[\frac{1}{2}
\left(v_{1}^{i''} - \vec{v}_1'' \cdot \vec{n}_1'' n_{1}^{i''} \right) 
- 4 v_{1}^{i''} \right] \nonumber \\ 
&&+\frac{m_2}{r_2''} \left[\frac{1}{2}
\left(v_{2}^{i''} - \vec{v}_2'' \cdot \vec{n}_2'' n_{2}^{i''} \right) 
- 4 v_{2}^{i''} \right] .
\ea
These expressions are accurate up to errors of $O(m/r_A)^2$. The 3-metric is
conformally flat and in the vicinity of each black hole (near $r_A=0$) the
metric is very similar to the metric of a black hole in isotropic
coordinates. Note that Ref.~\cite{Schafer:1986rd} uses units
where $16 \pi G=1$, but in Eq.~(\ref{admtt-metric}) we use units
where $G=1$ instead.

Also note that the PN expressions above are not pure Taylor expansions
in $m/r$ or $v/c$. Instead, certain higher order PN terms have been
added to make the metric similar to the Schwarzschild metric in
isotropic coordinates near each black hole. This resummation formally 
does not change the PN accuracy of the metric, but in practice
it improves the PN metric since it yields a metric that is very
similar to the inner zone black hole metric. The PN metric written in this
form above even has apparent horizons near $r_A=m_A/2$. Later when we
plot our results we will use this resummed form of the metric.
The reader may worry that there is a certain arbitrariness in adding
higher order terms. However, we have used the following two criteria 
to minimize this arbitrariness:
i) We only add terms that would actually be present at higher PN order 
(compare~\cite{Schafer:1986rd,Jaranowski97,Jaranowski:1997ky,Tichy:2002ec}).
ii) The terms added have coefficients of order unity, and
thus do not affect the usual PN order counting.
Notice that point i) of course only means that we have added some higher
order terms, not all higher order terms.

The lapse and the spatial metric are similar to what was used in paper $1$,
when we expand them in the buffer zone. However, the shift given here is
different from that in paper $1$ and, thus, the maps returned by asymptotic
matching will also be different.

Now that we have the $4$-metric in the near zone it is convenient to
remove the frame rotation by performing the transformation 
\ba
\label{corotatingtransformation}
t''&=&t,
\nonumber \\
x''&=&x\cos{\omega t} - y \sin{\omega t},
\nonumber \\
y''&=&x \sin{\omega t} + y \cos{\omega t},
\nonumber \\
z''&=&z,
\ea
where unprimed symbols stand for corotating ADMTT coordinates and
double primed symbols correspond to the inertial ADMTT frame.  After
doing this coordinate change, the form of the near zone metric
(\ref{admtt-metric}) as well as the lapse and conformal factor remain
unchanged, with only the shift picking up an additional $\omega \times
r$ term and now becomes
\ba
\beta^{(3) i} &=& 
\frac{m_1}{r_1} \left[\frac{1}{2}
\left(v_{1}^{i} - \vec{v}_1 \cdot \vec{n}_1 n_{1}^{i} \right) 
- 4 v_{1}^{i} \right]  \nonumber \\
&&+\frac{m_2}{r_2} \left[\frac{1}{2}
\left(v_{2}^{i} - \vec{v}_2 \cdot \vec{n}_2 n_{2}^{i} \right) 
- 4 v_{2}^{i} \right] \nonumber \\
&&- \epsilon_{ik3} \omega x^k ,
\ea
where $\vec{v}_A$ and $\vec{n}_A$ are now time independent
and equal to the double primed versions at $t=0$.
In these coordinates, the magnitude of the radial vectors become
\be 
r_A'' = r_A = \left(x_A^2 + y^2 + z^2\right)^{1/2}, 
\ee 
where $x_1 = x - b m_2/m$ and $x_2 = x + b m_1/m$.

Let us now make one final coordinate transformation where we will
shift the origin to the center of BH~1, i.e.,
\ba
\label{comshift}
x' &=& x_1 = x - \frac{m_2 b}{m},
\nonumber \\
y' &=& y, \qquad z' = z, \qquad t' = t,
\ea
and $r_A' = r_A$. The single-primed coordinates then stand for the
shifted corotating ADMTT coordinate system. This coordinate change is
useful because we will later concentrate on matching in ${\cal
  O}_{13}$. We will usually make explicit reference to $x_1$ instead
of $x'$ to remind us that this distance should be measured from BH~1.
This shift of the origin was not performed in paper $1$ and, thus, the
coordinate transformations that we will find will look slightly
different.  However, to $O(m/b)$ this transformation should reduce to
the results of paper $1$ when the shift is undone.

We now concentrate on the overlap region (buffer zone)
${\cal{O}}_{13}$.  Inside ${\cal O}_{13}$ we can expand all terms
proportional to $1/r_2$ in Legendre polynomials of the form
\be
\label{Legendre-exp}
\frac{1}{r_2'} = \frac{1}{b} \sum_{n=0}^{\infty} 
\left(-\frac{r_1'}{b}\right)^{n} P_n\left(\frac{x_1}{r_1'}\right). 
\ee
Substituting Eq.~(\ref{Legendre-exp}) into Eq.~(\ref{admtt-metric}),
we obtain
\begin{widetext}
\ba
\tilde{g}_{0'0'}^{(3)} &\sim& -1 + \frac{2m_1}{r_1'} + \frac{2m_2}{b}
\left[ 1 + \frac{r_1'}{b} P_1\left(\frac{x_1}{r_1'}\right) +
\left(\frac{r_1'}{b}\right)^2 P_2\left(\frac{x_1}{r_1'}\right) \right] +
\omega^2 \left[\left(x_1 + \frac{m_2 b}{m}\right)^2 + y^{'2}\right],
\nonumber \\
\tilde{g}_{0'1'}^{(3)} &\sim& - y' \; \omega \left\{ 1 + \frac{2
m_1}{r_1'} + \frac{2 m_2}{b} \left[1 +\frac{r_1'}{b}
P_1\left(\frac{x_1}{r_1'}\right)\right]\right\} - \frac{1}{2}
\frac{\mu}{r_1'} b \omega \frac{y' \left(x_1+\frac{m_2
b}{m}\right)}{r_1^{'2}} \left(1 - \frac{r_1^{'3}}{b^3} \right)
\nonumber \\ 
&& + \frac{\mu}{2 m} b \omega \frac{by}{r_1^{'2}} \left[\frac{m_2}{r_1'} +
\frac{m_1}{b} \frac{r_1^{'2}}{b^2} \left(1 + 3 \frac{r_1'}{b}
P_1\left(\frac{x_1}{r_1'}\right)\right)\right],
\nonumber \\
\tilde{g}_{0'2'}^{(3)} &\sim& \left(x_1+\frac{m_2 b}{m}\right) \;
\omega \left\{ 1 + \frac{2 m_1}{r_1'} + \frac{2 m_2}{b}
\left[1+{r_1'\over{b}}P_1\left(\frac{x_1}{r_1'} \right)\right]\right\} -
\frac{7}{2} \frac{\mu}{r_1'} b \omega \left\{ 1 -
{r_1'\over{b}}\left[1+{r_1'\over{b}}P_1\left(\frac{x_1}{r_1'}\right)
\right. \right. 
\nonumber \\
&&  \left. \left. +
\left(r_1'\over{b}\right)^2 P_2\left(\frac{x_1}{r_1'}\right)
\right]\right\} - \frac{1}{2} \frac{\mu}{r_1'} b \omega
\frac{y^2}{r_1^{'2}} \left(1 -\frac{r_1^{'3}}{b^3}\right), 
\nonumber \\ 
\tilde{g}_{0'3'}^{(3)} &\sim& - \frac{1}{2} \frac{\mu}{r_1'} b \omega
\frac{y z}{r_1^{'2}} \left(1 - \frac{r_1'}{b^3} \right),
\nonumber \\
\label{nearasympt}
\tilde{g}_{i'j'}^{(3)} &\sim& \delta_{ij} \left\{1 + {2m_1\over{r_1'}}
+ {2m_2\over{b}} \left[ 1 + {r_1'\over{b}}
P_1\left(\frac{x_1}{r_1'}\right) + \left(r_1'\over{b}\right)^2
P_2\left(\frac{x_1}{r_1'}\right) \right] \right\}, 
\ea
\end{widetext}
where all errors are of order $O(2,3)$ and where $m_1 \ll r_1 \ll b$.
Eq.~(\ref{nearasympt}) is denoted with a tilde because it is the
asymptotic expansion in the buffer zone around BH~1 of the ADMTT
metric. This metric has now been expressed entirely in terms of the
shifted corotating ADMTT coordinate system $(x',y',z',t')$.

Note that Eq.~(\ref{nearasympt}) is a bivariate series, since it
depends on two {\textit{independent}} expansion parameters $m_1/r_1'$
and $r_1'/b$. Since both expansion parameters must be
small independently, Eq.(\ref{nearasympt}) is valid in a $4$-volume
defined by the size of the buffer zone.

\section{Inner Zone: A Black Hole Perturbative Metric}
\label{inner}
In this section we discuss the metric in the inner zone ${\cal{C}}_1$
of BH~1 and its asymptotic expansion in the overlap region
${\cal{O}}_{13}$. Since this metric will be the same as that used in
the inner zone in paper $1$, we will minimize its discussion and
mostly refer to that paper. However, since we are using different
notation than that used in paper $1$, we will summarize here the
principal formulas.

In the inner zone $1$, let us use inertial isotropic coordinates,
labeled by $x^{\bar{i'}}=(\bar{x}',\bar{y}',\bar{z}',\bar{t}')$, which
are centered at BH $1$. Note that these coordinate are identical to
the shifted corotated harmonic coordinates used in the near zone to
zeroth order. Let us further denote the inner zone metric by
$g^{(1)}_{\bar{\mu}'\bar{\nu}'}$, which will be given by a tidally
perturbed Schwarzschild solution as
\begin{widetext}
\ba
\label{internalmetricIC}
g_{\bar{0}'\bar{0}'}^{(1)} &\approx& -\left[{1-{M_1/(2
      \bar{r}_1')}}\over{1+{M_1/(2 \bar{r}_1')}}\right]^2 +
{m_2\over{b^3}} \left(1-{M_1\over{2 \bar{r}_1'}}\right)^4 \left[ 3
  \left( \bar{x}' \cos{\Omega \bar{t}'} + \bar{y}' \sin{\Omega
      \bar{t}'} \right)^2 - \bar{r}_1'^2 \right],
\nonumber \\
g_{\bar{0}'\bar{1}'}^{(1)} &\approx& {2 m_2\over{b^3}}
\sqrt{m\over{b}} \left(1-{M_1\over{2\bar{r}_1'}}\right)^2
\left(1+{M_1\over{2\bar{r}_1'}}\right)^4
\left[\left(\bar{z}'^2-\bar{y}'^2\right)\sin{\Omega \bar{t}'} -
  \bar{x}'\bar{y}' \cos{\Omega \bar{t}'}\right],
\nonumber \\
g_{\bar{0}'\bar{2}'}^{(1)} &\approx& {2 m_2\over{b^3}}
\sqrt{m\over{b}} \left(1-{M_1\over{2\bar{r}_1'}}\right)^2
\left(1+{M_1\over{2\bar{r}_1'}}\right)^4
\left[\left(\bar{x}'^2-\bar{z}'^2\right)\cos{\Omega \bar{t}'} +
  \bar{x}'\bar{y}' \sin{\Omega \bar{t}'}\right],
\nonumber \\
g_{\bar{0}'\bar{3}'}^{(1)} &\approx& {2 m_2\over{b^3}}
\sqrt{m\over{b}} \left(1-{M_1\over{2\bar{r}_1'}}\right)^2
\left(1+{M_1\over{2\bar{r}_1'}}\right)^4 \left(\bar{y}' \cos{\Omega
    \bar{t}'} - \bar{x}' \sin{\Omega \bar{t}'}\right) \bar{z}',
\nonumber \\
g_{\bar{i}'\bar{j}'}^{(1)} &\approx& \left(1+{M_1\over{2
      \bar{r}_1'}}\right)^4 \left( \delta_{ij} + {m_2\over{b^3}}
  \left[3 \left(\bar{x}' \cos{\Omega \bar{t}'} + \bar{y}' \sin{\Omega
        \bar{t}'}\right)^2 - \bar{r}_1'^2 \right] \left\{ \left[
      \left(1+{M_1\over{2 \bar{r}_1'}}\right)^4 - {2
        M_1^2\over{\bar{r}_1'^2}}\right] \delta_{ij} \right. \right.
\nonumber \\
&& \left. \left. - {2 M_1\over{\bar{r}_1'}} \left(1 + {M_1^2\over{4
          \bar{r}_1'^2}}\right) {\bar{x}^{i'}
      \bar{x}^{j'}\over{\bar{r}_1'^2}} \right\} \right),
\ea
\end{widetext}
where $\bar{r}'_1 = (\bar{x}'^2 + \bar{y}'^2 + \bar{z}'^2)^{1/2}$.
This equation is identical to that used in the inner zone of paper $1$
or Eq.~($3.23$) of Ref.~\cite{Alvi:1999cw}. This metric was first
computed by Alvi~\cite{Alvi:1999cw} by linearly superposing a
Schwarzschild metric to a tidal perturbation near BH $1$. This
perturbation is computed as an expansion in $\epsilon_{1,2} =
\bar{r}'_{1,2}/b$ and it represents the tidal effects of the external
universe on BH $1$. Since a simple linear superposition would not
solve the Einstein equations, multiplicative scalar functions are
introduced into the metric perturbation, which in turn are determined
by solving the linearized Einstein equations. With these scalar
functions, the metric then solves the linearized Einstein equations
and it represents a tidally perturbed Schwarzschild black hole. This
metrics has been found to be isomorphic to that computed by
Poisson~\cite{Poisson:2005pi} in advanced Eddington-Finkelstein
coordinates.

Asymptotic matching will be easier when performed between metrics in
similar coordinate systems. We, thus, choose to make a coordinate
transformation to corotating isotropic coordinates $x^{\bar{i}} =
(\bar{x},\bar{y},\bar{z},\bar{t})$. The inner zone metric in
corotating isotropic coordinates will be denoted by
$g^{(1)}_{\bar{\mu}\bar{\nu}}$ and is given by
\ba
\label{internalmetricICC}
g^{(1)}_{\bar{0}\bar{0}} &\approx& H_t + H_{s1} \Omega^2
\left(\bar{x}^2+\bar{y}^2\right) + 2 H_{st} \bar{x} {\Omega\over{b^2}}
\left(\bar{x}^2+\bar{y}^2-\bar{z}^2\right), 
\nonumber \\
g^{(1)}_{\bar{0}\bar{i}} &\approx&  - H_{s1} \Omega
\epsilon_{\bar{i}\bar{j}\bar{3}} x^{\bar{j}} + \frac{H_{st}}{b^2}
\left[ \bar{y} \left(\delta_{\bar{i}}^{\bar{3}} \bar{z} -
    \delta_{\bar{i}}^{\bar{1}} \bar{x} \right) + \left(\bar{x}^2 - \bar{z}^2\right)
  \delta_{\bar{i}}^{\bar{2}} \right], 
\nonumber \\ 
g^{(1)}_{\bar{i}\bar{j}} &\approx& \delta_{\bar{i}\bar{j}} H_{s1} -
H_{s2} \frac{x^{\bar{i}} x^{\bar{j}}}{b^2},  
\ea
where $\epsilon_{\bar{i}\bar{j}\bar{k}}$ is the standard Levi-Civita
symbol with convention $\epsilon_{\bar{1}\bar{2}\bar{3}} = 1$ and
where $\delta^{\bar{a}}_{\bar{b}}$ is the Kronecker delta. In
Eq.~(\ref{internalmetricICC}) we use the shorthand
\begin{widetext}
\ba
H_{st} &=& 2 m_2 \sqrt{\frac{m}{b^3}} 
\left(1 - {M_1\over{2 \bar{r}_1}}\right)^2 \left(1 + {M_1\over{2 \bar{r}_1}} \right)^4,
\nonumber \\ 
H_{s1} &=& \left(1 + {M_1\over{2 \bar{r}_1}}\right)^4 \left\{1 + 2{m_2\over{b^3}} 
\bar{r}_1^2 P_2\left({\bar{x}\over{\bar{r}_1}}\right)
\left[\left(1 + {M_1\over{2 \bar{r}_1}}\right)^4 - 2 {M_1^2\over{\bar{r}_1^2}}\right]\right\}, 
\nonumber \\
H_{s2} &=& \left(1+{M_1\over{2\bar{r}_1}}\right)^4 \left(1+{M_1^2\over{4\bar{r}_1^2}}\right) {4m_2M_1\over{b \bar{r}_1}} 
P_2\left({\bar{x}\over{\bar{r}_1}}\right) ,
\nonumber \\ 
H_{t} &=& - \left({1-M_1/2 \bar{r}_1}\over{1+M_1/2 \bar{r}_1}\right)^2 + 2 \left(1-{M_1\over{2 \bar{r}_1}}\right)^4
{m_2\over{b^3}} \bar{r}_1^2 P_2\left({\bar{x}\over{\bar{r}_1}}\right),
\ea
\end{widetext}
where $\bar{r}_1 = \bar{r}_1'$ and the errors are still of
$O(\bar{r}_1/b)^3$. These shorthands for the different components of
the metric are identical to those used in paper $1$.

We now need to asymptotically expand the inner zone metric in the
buffer zone, {\textit{i.e.}}, as $M_1/\bar{r}_1 \ll 1$. Doing so we obtain 
\narrowtext{
\ba
\label{internalmetricAICC}
\tilde{g}_{\bar{0}\bar{0}}^{(1)} &\sim&  -1 + {2 M_1\over{\bar{r}_1}} + {2
  m_2\over{b^3}} \bar{r}_1^2 P_2 \left(\bar{x}\over{\bar{r}_1}\right)
\nonumber \\
&& + \Omega^2 \left(\bar{x}^2+\bar{y}^2\right), 
\nonumber \\ 
\tilde{g}_{\bar{0}\bar{1}}^{(1)} &\sim& {-2 m_2\over{b^3}}
  \sqrt{\frac{m}{b}} \bar{y} \bar{x} - \bar{y} \Omega 
\left(1 + {2 M_1\over{\bar{r}_1}} \right),
\nonumber \\
\tilde{g}_{\bar{0}\bar{2}}^{(1)} &\sim& {2 m_2\over{b^3}}
  \sqrt{\frac{m}{b}} \left(\bar{x}^2-\bar{z}^2\right)  
+ \bar{x} \Omega \left(1 + {2 M_1\over{\bar{r}_1}}\right),
\nonumber \\ 
\tilde{g}_{\bar{0}\bar{3}}^{(1)} &\sim& {2 m_2\over{b^3}}
  \sqrt{\frac{m}{b}} \bar{z} \bar{y},  
\nonumber \\ 
\tilde{g}_{\bar{i}\bar{j}}^{(1)} &\sim& \delta_{ij} \left[ 1 + {2
  M_1\over{\bar{r}_1}} + {2 m_2\over{b^3}} \bar{r}_1^2
  P_2\left(\bar{x}\over{\bar{r}_1}\right) \right] 
\ea
}
The asymptotic expansion of $g^{(1)}_{\bar{\mu}\bar{\nu}}$ will be
denoted by $\tilde{g}^{(1)}_{\bar{\mu}\bar{\nu}}$. Note that this
asymptotic expansion is a bivariate expansion in both $\bar{r}_1/b$
and $M_1/\bar{r}_1$. In other words, it is the asymptotic expansion in
the buffer zone to the approximate solution in the inner zone.

\section{Matching Conditions and Coordinate Transformations}
\label{matching}
In this section, we find the coordinate and parameter maps that relate
adjacent solutions. Since the coordinate systems are similar to each
other in the buffer zone, we make the following ansatz for the
transformation between coordinates and parameters
\begin{widetext}
\ba
\label{coordtransf}
\bar{x} &\approx& x' \left[ 1 + 
\left(m_2\over{b}\right)^{1/2} \chi_1(x^{\mu'}) + 
\left(m_2\over{b}\right) \chi_2(x^{\mu'})
+\left(m_2\over{b}\right)^{3/2} \chi_3(x^{\mu'}) \right],
\nonumber \\
\bar{y} &\approx& y' \left[ 1 + \left(m_2\over{b}\right)^{1/2}  \gamma_1(x^{\mu'}) 
+ \left(m_2\over{b}\right) \gamma_2(x^{\mu'}) 
+ \left(m_2\over{b}\right)^{3/2} \gamma_3(x^{\mu'}) \right],
\nonumber \\
\bar{z} &\approx& z' \left[ 1 + \left(m_2\over{b}\right)^{1/2} \zeta_1(x^{\mu'}) 
+ \left(m_2\over{b}\right) \zeta_2(x^{\mu'}) 
+ \left(m_2\over{b}\right)^{3/2} \zeta_3(x^{\mu'}) \right],
\nonumber \\
\bar{t} &\approx& t' \left[ 1 + \left(m_2\over{b}\right)^{1/2}  \tau_1(x^{\mu'}) + 
\left(m_2\over{b}\right) \tau_2(x^{\mu'}) 
+ \left(m_2\over{b}\right)^{3/2} \tau_3(x^{\mu'}) \right],
\nonumber \\
M_1 &\approx& m_1 \left[1 + \left(m_2\over{b}\right)^{1/2} \eta_1 + {m_2\over{b}} \eta_2  
+ \left(m_2\over{b}\right)^{3/2} \eta_3 \right],
\nonumber \\
\Omega &\approx& \omega \left[1 + \left(m_2\over{b}\right)^{1/2} \kappa_1 
+ {m_2\over{b}} \kappa_2 + \left(m_2\over{b}\right)^{3/2} \kappa_3 \right],
\ea
\end{widetext}
where, as in paper $1$, $\chi_{1,2,3}$, $\gamma_{1,2,3}$, 
$\zeta_{1,2,3}$ and $\tau_{1,2,3}$ are functions of the coordinates,
whereas $\eta_{1,2,3}$ and $\kappa_{1,2,3}$ are
coordinate independent. Note that this ansatz is slightly different
from the one made in paper $1$, because here we have shifted the
origin of the near zone coordinate system, so that to zeroth order it
agrees with the coordinates used in the inner zone. In order to
determine these maps, we enforce the matching condition,
{\textit{i.e.}}
\be
\label{ourmatch}
\tilde{g}_{\mu'\nu'}^{(3)}(x^{\lambda'}) \sim 
\tilde{g}_{\bar{\rho} \bar{\sigma}}^{(1)}\left(\bar{x}^{\lambda}\left(x^{\lambda'}\right)\right)
 {{\partial \bar{x}^{\rho}}\over{\partial x^{\mu'}}}
{{\partial \bar{x}^{\sigma}}\over{\partial x^{\nu'}}},
\ee
which leads to a system of first-order coupled partial differential
equations, which we solve order by order.

The system to $O(m/b)^0$ provides no information because both metrics
are asymptotic to Minkowski spacetime to lowest order. The first
non-trivial matching, occurs at $O(m/b)^{1/2}$. The differential
system at this order and at order $O(m/b)$ are similar to those obtain
in paper $1$ and we, thus, omit them here.  The solution up to
$O(m/b)$ is given by
\ba
\label{match:1PN}
\bar{x} &\approx& x' \left[ 1 + \frac{m_2}{b} \left( 1 -
    \frac{x'}{2b} \right) + \left(\frac{m_2}{b}\right)^{3/2}
  \chi_3(x^{\mu'}) \right] 
\nonumber \\
&& + {m_2\over{2 b^2}} \left( 2 t'^2 + y'^2 +z'^2 \right), 
\nonumber \\
\bar{y} &\approx& y' \left[ 1 + \left(m_2\over{b}\right) \left(1 -
    \frac{x'}{b} \right)  + \left(m_2\over{b}\right)^{3/2}
  \gamma_3(x^{\mu'}) \right] 
\nonumber \\
&& + \sqrt{\frac{m_2}{b}} \sqrt{\frac{m_2}{m}} t', 
\nonumber \\
\bar{z} &\approx& z' \left[ 1 + \left(m_2\over{b}\right) \left ( 1 -
    \frac{x'}{b} \right) + \left(m_2\over{b}\right)^{3/2} \zeta_3(x^{\mu'})
  \right], 
\nonumber \\
\bar{t} &\approx& t' \left[ 1 - \left(m_2\over{b}\right) \left( 1 -
    \frac{x'}{b} \right) + \left(m_2\over{b}\right)^{3/2}
  \tau_3(x^{\mu'}) \right], 
\ea
with $\eta_1=0$ and $\kappa_1=0$. We should note that in solving the
differential system to $O(m/b)^{1/2}$ and $O(m/b)$ we have explicitly
required that the coordinate system be not boosted or rotating with
respect to each other. Furthermore, we found the particular solution
where the $t'=0$ and the $\bar{t}=0$ slices coincide, since $t'=0$
implies $\bar{t}=0$. We should also note that Eq.~(\ref{match:1PN}) of
this paper is different from the transformation found in paper $1$
because in this paper we have shifted the origin of the near zone
coordinate system, which simplifies the transformation.

Matching at $O(m/b)^{3/2}$ needs to be performed only on the $0i$
components of the metric to get the first non-trivial correction to
the extrinsic curvature. This order counting does not follow the
standard post-Newtonian scheme and it is explained in more detail in
paper $1$.  Since the near zone shift is different from that used in
paper $1$, the differential system obtained by imposing the matching
condition of Eq.~(\ref{ourmatch}) is also different. This system is
given by
\begin{widetext}
\ba
\label{pdeND:3/2}
t' \tau_{3,x'} - x' \chi_{3,t'} &\sim& \sqrt{\frac{m}{m_2}} \left\{ \frac{y'x'}{b^2}
\left[\frac{m_1}{2 m} \left(\frac{b^3}{r_1^{'3}} - 1
    + 3 \frac{m_1}{m} \right) - 4 \right] + \frac{y'}{b} \left[
 - \frac{m_2}{m} - \kappa_2 + 1 - \frac{m_1}{2 m} \right] \right\}, 
\nonumber \\
t' \tau_{3,y'} - y' \gamma_{3,t'} &\sim& \sqrt{\frac{m}{m_2}}
\left\{\frac{z'^2}{b^2} \left(\frac{7 m_1}{4 m} - \frac{3}{2} - \frac{m_2}{m}\right) +
  \frac{y'^2}{b^2} \left[ \frac{m_1}{2m} \left( \frac{b^3}{r_1'^3} +
      \frac{5}{2} \right) - \frac{m_2}{m} -
  \frac{1}{2}\right] + \frac{t'^2}{b^2} + \frac{x'^2}{b^2} \left( \frac{7}{2} + 2
  \frac{m_2}{m} 
\right. \right.
\nonumber \\
&& \left. \left.  - \frac{7 m_1}{2 m} \right)+ \frac{x'}{b} \left(\kappa_2 -1
  + \frac{m_2}{m} + \frac{7 m_1}{2 m} \right) 
+ \frac{3}{2} - \frac{m_2}{m} + \frac{7 m_1}{2 m} \left( -1 +
  \frac{b}{r_1'} \right) - \frac{\mu}{2 m} \right\},
\nonumber \\
t' \tau_{3,z'} - z' \zeta_{3,t'} &\sim& \sqrt{\frac{m}{m_2}} \frac{y'z'}{b^2}
\left[\frac{m_1}{2 m} \left(\frac{b^3}{r_1'^3} -1 \right) + 1 \right].
\ea
\end{widetext}
Note that $\kappa_2$ represents the degree of freedom associated with
the angular velocity.
For simplicity, we now set $\kappa_2 = 0$.
Also note that $\eta_2$, $\eta_3$ and $\kappa_3$ never enter the
matching conditions to this order, so we set those quantities to zero.
All the matching relations between the parameters local to different
approximation have now been specified to $O(m/b)^2$, {\textit{i.e.}}
\be
\label{matching-conds}
\Omega \approx \omega, 
\quad M_1 \approx  m_1.
\ee

With these matching relations, we can now find a particular solution
to Eqs.~(\ref{pdeND:3/2}) and this set will produce matching. Once
more, we choose constants of integration such that the $t'=0$ and the
$\bar{t}=0$ slices coincide. Doing so we obtain the solution
\begin{widetext}
\ba
\label{fulltransf-nokappa}
\bar{x} &\approx& x_1 \left( 1 + \frac{m_2}{b} \left( 1 -
    \frac{x_1}{2b} \right) - \left(\frac{m_2}{b}\right)^{3/2} 
  \sqrt{\frac{m}{m_2}} \left\{ \frac{y t}{b^2} \left[\frac{m_1}{2 m}
    \left(\frac{b^3}{\underline{r}_1^3} - 1  + \frac{3 m_1}{m}  \right) -4
    \right] + \frac{y}{b} \left( 1 - \frac{m_2}{m} - \frac{m_1}{2 m}
    \right) \right\} \right) 
\nonumber \\
&& + {m_2\over{2 b^2}} \left( 2 t^2 + y^2 +z^2 \right),  
\nonumber \\
\bar{y} &\approx& y \left[ 1 + \left(m_2\over{b}\right) \left(1 -
    \frac{x_1}{b} \right) \right]  - \left(m_2\over{b}\right)^{3/2}
 t \sqrt{\frac{m}{m_2}} \left\{\frac{z^2}{b^2} \left(\frac{7 m_1}{4
    m} - \frac{3}{2} - \frac{m_2}{m}\right) + \frac{y^2}{b^2} \left[
    \frac{m_1}{2m} \left( \frac{b^3}{\underline{r}_1^3} + \frac{5}{2} \right) - \frac{m_2}{m} -
  \frac{1}{2}\right] + \frac{t^2}{3 b^2} \right.
\nonumber \\
&& \left. + \frac{x_1^2}{b^2} \left( \frac{7}{2} + 2 \frac{m_2}{m} - \frac{7
    m_1}{2 m} \right) + \frac{x_1}{b} \left( -1 + \frac{m_2}{m} +
    \frac{7 m_1}{2m} \right)  + \frac{3}{2} - \frac{m_2}{m}
  + \frac{7 m_1}{2 m} \left( -1 + \frac{b}{\underline{r}_1} \right) - \frac{\mu}{2
    m} \right\} + \sqrt{\frac{m_2}{b}} \sqrt{\frac{m_2}{m}} t, 
\nonumber \\
\bar{z} &\approx& z \left\{ 1 + \left(m_2\over{b}\right) \left ( 1 -
    \frac{x_1}{b} \right) - \left(m_2\over{b}\right)^{3/2} 
 \sqrt{\frac{m}{m_2}} \frac{y t}{b^2} \left[\frac{m_1}{2 m}
    \left(\frac{b^3}{\underline{r}_1^3} -1 \right) + 1 \right] \right\}, 
\nonumber \\
\bar{t} &\approx& t \left[ 1 - \left(m_2\over{b}\right) \left( 1 -
    \frac{x_1}{b} \right) \right],
\ea
\end{widetext}
where errors are of $O(2,3)$. 

As in paper $1$, this coordinate transformation is singular at
$r_1'=0$.  Note, however, that in Eq.~(\ref{fulltransf-nokappa}) we
have replaced $r_1'$ by 
\be
\label{r-shift}
\underline{r}_{1} = \sqrt{r_1'^2 + 6 m^2}.
\ee 
This replacement amounts to adding higher order terms to the coordinate
transformation, which have no effect in the buffer zone at the current level
of accuracy. Yet it has the advantage that the resulting coordinate
transformation is now regular at $r_1'=0$. Thus, the replacement does not
affect asymptotic matching (in the buffer zone). It merely leads to an inner
zone metric expressed in a better coordinate system.

In paper $1$ there was one term in the coordinate transformation that
was singular at the location of the holes, whereas here there are $3$
such singular terms. This increase in singular behavior is due to the
ADMTT shift having more poles than the shift of paper $1$. The shift
of the origin into the complex plane, as introduced in
Eq.~(\ref{r-shift}), removes this singular behavior from the real line,
while only introducing uncontrolled remainders at a higher order.  We
should also note that in Eq.~(\ref{fulltransf-nokappa}), we undid the
shift given by Eq.~(\ref{comshift}), so that we measure distances from
the center of mass of the system.

The coordinate transformations presented in
Eq.~(\ref{fulltransf-nokappa}) is only valid in ${\cal{O}}_{13}$, but
we can find the transformation in ${\cal{O}}_{23}$ by a simple
symmetry transformation. This transformation is given by the
following rules: substitute $1\leftrightarrow2$ and
\be
\label{symmetry}
x \to -x, \qquad    y \to -y, \qquad z \to z.
\ee

The set of matching conditions plus the coordinate transformation
presented in Eqs.~(\ref{matching-conds})
and~(\ref{fulltransf-nokappa}) are the end result of the matching
scheme. 
Now that we have an approximate global metric, we can
decompose it to provide initial data for numerical simulations.  This
decomposition consists of a foliation of the manifold with spacelike
hypersurfaces, where we choose the slicing given by $t=0=\bar{t}$.
Notice that we have found a coordinate transformation that is
consistent both with this foliation and with the condition $m_A \ll
r_A \ll b$, valid in the buffer zone.

Note that there is some freedom in the matching scheme at the order of
accuracy of this paper. This freedom is rooted in that a different
choice for the matching parameters ($M$ and $\Omega$) could be
made. However, except for the $\kappa_2$ term in $\Omega$,
this would have no effect on the coordinate transformation or the
$4$-metric at the orders considered here. Furthermore, one
can show that the value of $\kappa_2$ does not affect the physics of the
initial data we will construct below. Above, the choice of 
$\kappa_2 = 0$ was made in order to simplify the relation
between the parameters of the two approximations. Specifically, it was chosen
such that the angular velocity parameters of the two approximations are
equal. This is, however, not the only possible choice. Fortunately, the
choice of $\kappa_2$ will not affect the initial data we will construct
later, since both the 3-metric and the extrinsic curvature will change only
by uncontrolled remainders of higher order. This happens for the 
following reasons. A non-zero $\kappa_2$ would only change the
spatial part of the coordinate transformation at $O(m/b)^{3/2}$,
and thus would have no effect on the slicing.
Since the leading order non-zero piece in the extrinsic
curvature is already of $O(m/b)^{3/2}$ and proportional to $\Omega$,
any change due to spatial coordinate transformations or parameter changes at
$O(m/b)^{3/2}$ will cause changes only at even higher order. For the
$3$-metric, coordinate transformations or parameter changes 
at $O(m/b)^{3/2}$ will make a difference at order $O(m/b)^{3/2}$ 
in the metric. We can, however, again neglect these changes,
since the $3$-metric was only matched up to $O(m/b)$.
In conclusion, this means that any allowed change in the matching parameters
will not affect the physics of the initial data sets constructed below.

\section{Constructing a Global Metric}
\label{secglobal}
In this section, we present the global metric, by performing
the coordinate transformation found in the previous section on the
inner zone metric. The piecewise metric in corotating ADMTT
coordinates is given by
\ba
\label{globalpiece}
g^{(global)}_{\mu\nu} &\approx&  \left\{  \begin{array} {ll}
                  g^{(1)}_{\mu\nu}, \qquad {\textrm{in}} \; {\cal{C}}_{1},
                  \\
                  g^{(2)}_{\mu\nu}, \qquad {\textrm{in}} \; {\cal{C}}_{2},
                  \\
                  g_{\mu\nu}^{(3)}, \qquad {\textrm{in}} \; {\cal{C}}_{3}.
\end{array} \right.  
\ea
In Eq.~(\ref{globalpiece}), $g^{(3)}_{\mu\nu}$ denotes the near zone
metric given in Eq.~(\ref{admtt-metric}), but in unprimed
corotating ADMTT coordinates, whereas $g^{(1)}_{\mu\nu}$ and $g^{(2)}_{\mu\nu}$
stand for the inner zone metrics of BH $1$ and $2$ transformed to
corotating ADMTT coordinates via Eq.~(\ref{fulltransf-nokappa}). Explicitly,
the inner zone metric is given by
\be
\label{innerzonemetric-transf}
g^{(1,2)}_{\mu\nu}
= g^{(1,2)}_{\bar{\mu}\bar{\nu}} J^{\bar{\mu}}_{\mu} J^{\bar{\nu}}_{\nu},
\ee
where $J^{\bar{\mu}}_{\nu}$ is the Jacobian of Eq.~(\ref{fulltransf-nokappa}), namely
\be
\label{jacobian1}
J^{\bar{\mu}}_{\nu} = \frac{ \partial x^{\bar{\mu}} }{ \partial x^{\nu} }.
\ee
In Eq.~(\ref{innerzonemetric-transf}), $g^{(1)}_{\bar{\mu}\bar{\nu}}$
is the inner zone metric in corotating isotropic coordinates given by
Eqs.~(\ref{internalmetricICC}), while $g^{(2)}_{\bar{\mu}\bar{\nu}}$
can be obtained by applying Eq.~(\ref{symmetry}) to
$g^{(1)}_{\bar{\mu}\bar{\nu}}$.  Eq.~(\ref{innerzonemetric-transf}) is
also expanded in paper $1$ with the substitution $h^{(1)}_{\mu\nu} =
g^{(1)}_{\bar{\mu}\bar{\nu}}$.  Note that the Jacobian given in
Appendix \ref{appendix_Jac} will be different from that presented 
in paper $1$ because the coordinate transformation is different.

In the following we present plots of different metric components at
time ${\bar{t}}=t'=0$. These plots are representative samples 
of the initial data and will show how well the matching procedure 
works. In most of these figures we choose to plot along
the axis that connects the holes (the $x$-axis). We have checked that
the behavior along other axis is qualitatively similar to that along 
the $x$-axis, as evidenced by the contour plots included in the section.
Furthermore, we plot only the relevant components because either the other
components vanish along the x-axis or they present similar behavior.
Thus the components and axis chosen present the general
behavior of the data.

For these plots, we choose two different physical systems. The first
system consists of equal mass black holes separated by $b=10m$.  The
two black holes are located at $x = \pm 5m$ and are surrounded by
apparent horizons, which approximately are coordinate spheres of
radius $m/4$. For this system, some quantities, like the
$xx$-component of the metric, are reflection symmetric under $x \to
-x$ and we then plot only the positive x-axis, omitting the inner
zone $2$ data.  The second system consists of two holes with mass
ratio $m_2/m_1=0.25$ separated by $b = 20 m$. Here the holes are located
at $x=4m$ and $x=-16m$, and surrounded by apparent horizons with radii 
of approximately $m_1/2$ and $m_2/2$ for holes $1$ and $2$
respectively. In general, a dashed line corresponds to
the post-Newtonian near zone metric, while the dotted or
dot-dot-dashed lines represent the perturbed Schwarzschild inner zone
metrics. 

Most figures will contain error bars, which in the case of the metric
are given by 
\ba
\label{metric_errors}
E_{near} &=& \frac{3}{2} \left(\frac{m_1^2}{r_1^2} +
  \frac{m_2^2}{r_2^2} \right) + \frac{5}{4} \frac{m_1
  m_2}{b} \left(\frac{1}{r_1} + \frac{1}{r_2} \right) 
\nonumber \\
&& - 2 \frac{m_1 m_2}{S^2},
\nonumber \\
E_{inner,A} &=& 2 \frac{m}{b} \left(\frac{r_A}{b}\right)^3,
\ea
where $S=r_1 + r_2 + b$. These error bars were obtained by analyzing
the next order term in the ADMTT PN metric and in the black hole
perturbation approximation. In particular, the PN error is different
from that used in paper $1$. Here the first term comes from the
expansion of the conformal factor and the last two from the $2$ PN
part of the transverse-traceless term in the ADMTT metric.  The full
PN error, given by the $3.5$ ADMTT PN metric, is much more complex
than the simplified approximation used in $E_{near}$, where the former
contains some terms that scale for example as $m_A^2/r_A^3$ and
$m_A^3/(S r_A^2)$. However, here we have neglected those terms and we
have checked that $E_{near}$ models well both the magnitude and
functional behavior of the full error in the regions plotted in all
figures with fractional errors of less than $1 \%$ everywhere.

\begin{figure}[t]
\includegraphics[scale=0.33,clip=true]{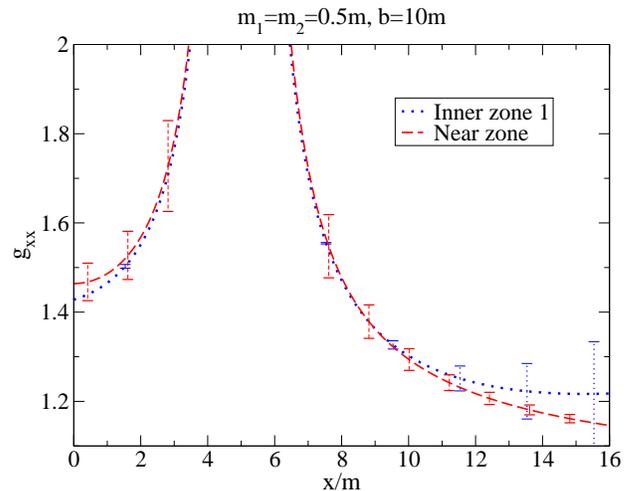}
\caption{\label{gxx} This figure shows the $xx$-component of the near
  zone (dashed line) and inner zone $1$ metric (dotted line) along the
  positive $x$-axis in ADMTT coordinates. The black holes are located
  at $x/m \approx \pm 5$ and the perturbative parameter is $m/b=1/10$.
  Observe that the metrics become asymptotic to each other in the
  buffer zone,  whose approximate center is roughly $4 m$ away from
  each 
  black hole. Also plotted are estimates of the error in the
  approximations.
  Note that the off-diagonal components of the 3-metric vanish
  along the x-axis for symmetry reasons, and that the $yy$- and 
  $zz$-components are very similar to the $xx$-component.
}
\end{figure}
In Fig.~\ref{gxx}, we plot the $xx$-component of the piece-wise
metric along the positive $x$-axis. Observe that the inner zone
metric of black hole $1$ (dotted line) is close to the post-Newtonian
near zone metric (dashed line) in the buffer zone of black hole $1$
given by $m_1 \ll \bar{r}_1 \ll b$. 
Recall that the definition of the buffer zone in
the theory of asymptotic matching is inherently imprecise in the sense
that as one approaches either the inner or outer radius of the buffer
zone shell, one of the two approximations (i.e. near zone or inner
zone metric) has much larger errors than the other. In our case we
need both $\bar{r}_A/b$ and $m_A/\bar{r}_A$ to be small, which will not be the
case at either end of the interval $m_A \ll \bar{r}_A \ll b$.  Hence we
expect the best agreement between the two approximations to occur
somewhere in the middle of each buffer zone shell. In fact the best
agreement should occur where each of the two approximations have
roughly the same error.  From Fig.~\ref{gxx} we see that the inner
zone metric of each black hole agrees with the near zone metric within
error bars in the middle of each of the two buffer zones, where both
have comparable errors. This agreement means that we have performed
successful matching. However, our result in Fig.~\ref{gxx} is even
better than this, since the near zone 3-metric is close to each of the
inner zone 3-metrics even near the black holes, where post-Newtonian
theory has large errors and is expected to fail.  The reason for this
somewhat surprising success near the black holes is that we have used
resummed post-Newtonian expressions
[Eqs.~(\ref{admtt-metric}),~(\ref{Psi_ADMTT}),~(\ref{alpha_ADMTT})
and~(\ref{beta_ADMTT})] in the ADMTT gauge, which are very close to
the isotropic Schwarzschild metric used in the inner zone.
Note that the off-diagonal components of the 3-metric vanish
along the x-axis for symmetry reasons, and that the $yy$- and 
$zz$-components are very similar to the $xx$ component.

\begin{figure}[t]
\includegraphics[scale=0.33,clip=true]{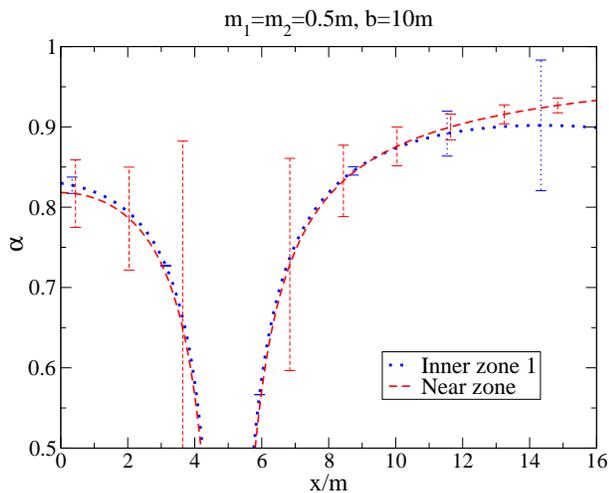}
\caption{\label{lapse}
  Plot of the lapse along the positive $x$-axis for the inner zone and
  near zone metrics. Note that the lapses are similar to each other
  not only in the buffer zone, but also near the hole. }
\end{figure}
\begin{figure}[t]
\includegraphics[scale=0.33,clip=true]{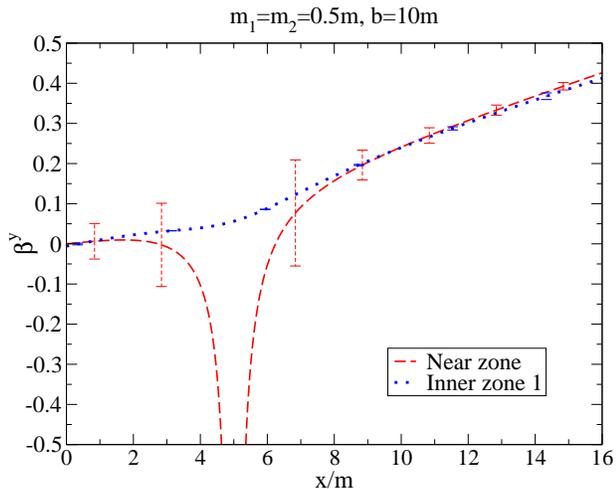}
\caption{\label{shift} 
  Plot of the $y$-component of the shift along the positive $x$-axis
  for the inner zone and near zone metrics.  Observe that although the
  inner zone and near zone curves are similar in the buffer zones, in
  this case they differ near the black holes. Also note that
  the $x$- and $z$-components of the shift vanish along the $x$-axis.}
\end{figure}
The quality of the asymptotic matching (in the buffer zone) is equally
good for all other components of the 4-metric.  This is evidenced in
Figs.~\ref{lapse} and \ref{shift} which show the near and inner zone
lapse and y-component of the shift, computed from the
4-metric~(\ref{globalpiece}) via
\ba
\beta^i_{} &=& g^{ij} g_{oj}.
\nonumber \\
\alpha_{} &=& = \left( g_{ok}\beta^k - g_{00} \right)^{1/2}.
\ea
As one can see from Fig.~\ref{lapse} the near (dashed line) and inner
zone $1$ lapse (dotted line) agree well in the buffer zone, located
about $4m$ away from each black hole.  Furthermore, the resummed
expression [Eq.~(\ref{alpha_ADMTT})] we use for the near zone lapse
also is very close to the inner zone lapse near each black hole, where
post-Newtonian theory has large error bars. In Fig.~\ref{shift} we see
that the near zone shift (dashed line) is close to the inner zone $1$
shift (dotted line) in the buffer zone as expected after matching
there. However, in the inner zone the post-Newtonian near zone shift
is not valid anymore and deviates strongly from the inner zone shift.

Thus in summary asymptotic matching works well for all components of
the 4-metric in the buffer zone. In addition, for most components
(except the shift) the resummed post-Newtonian expressions
[Eqs.~(\ref{admtt-metric}), (\ref{Psi_ADMTT}), (\ref{alpha_ADMTT}) and
(\ref{beta_ADMTT})] we use in the near zone, follow the inner zone
behavior even near the black holes, where this is not necessarily
expected.

\begin{figure}[t]
\includegraphics[scale=0.33,clip=true]{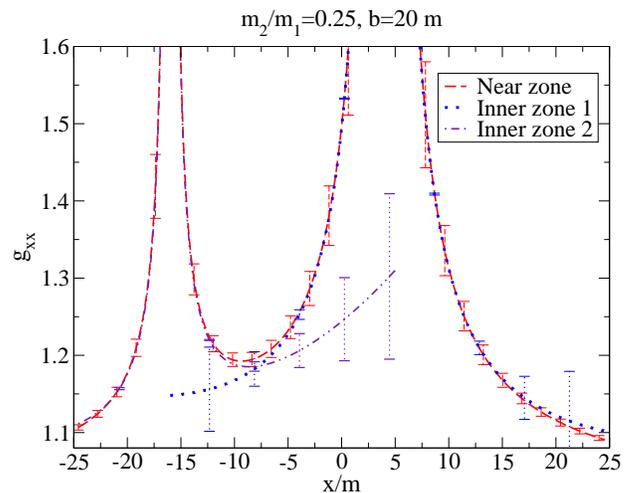}
\caption{\label{gxx-mu} This figure shows the $xx$ component of the
  piece-wise metric for a system of unequal masses and at larger
  separations. In particular, we have used $m_2/m_1 = 0.25$ and $b=20
  m$. Note that the matching improves since the perturbation 
  parameter $m/b$ is smaller.}
\end{figure}

We have also tested the coordinate and parameter maps for system with
unequal masses and found that matching is still valid.
Figure~\ref{gxx-mu} shows the $xx$-component of the piece-wise metric
along the entire $x$-axis for the case of $m_2/m_1 = 0.25$ and $b=20m$.
Observe that the matching improves in this case because the
perturbation parameter $m/b$ has become smaller. Also note that the
general features of the matching are the same as those presented in
the equal mass case. We have plotted along the entire x-axis and show
both the metrics of inner zone $1$ (dotted line) and $2$
(dot-dot-dashed line). 

Since we are interested in initial data on the ${\bar{t}}=t'=0$ slice
we also have to discuss the extrinsic curvature given by
\be
\label{K_ij_def}
K_{ij} = -\frac{1}{2\alpha} \left(\partial_t g_{ij} - \pounds_{\beta}
  g_{ij}\right).
\ee
In the inner zones we use this equation to numerically compute
$K_{ij}^{(1)}$ and $K_{ij}^{(2)}$.  Note however, that the inner zone
extrinsic curvature in corotating isotropic coordinates can also be
found in paper $1$, where $K^{(1), ICC}_{ij}=K^{(1)}_{\bar{i}\bar{j}}$
in the new notation.

The near zone extrinsic curvature can easily be computed analytically
from the near zone post-Newtonian 4-metric in ADMTT coordinates.
The result~\cite{Tichy:2002ec}
\ba
\label{ADMTT-curvature}
K_{ij}^{(3)}
&=& \Psi^{-2} \sum_{A=1}^2 \frac{3}{2 r_{A}^2}\left[ 
  p_A^i n_A^j +p_A^j n_A^i \right.
\nonumber \\
&& \quad
\left. -p_A^m n_A^n \delta_{mn}(\delta_{ij} -n_A^i n_A^j) 
  \right] .
\ea
is of Bowen-York form, and $v_{A}^{i}$ and
$n_{A}^{i}$ denote the particle velocities and directional vectors
already introduced in Sec.~\ref{near}. On the $t'=0$ slice these
quantities become
\be
v_{1}^{2}=  \omega \frac{m_2}{m} b ,\qquad
v_{2}^{2}= -\omega \frac{m_1}{m} b ,\qquad
v_{A}^{1}=v_{A}^{3}=0,
\ee
and
\ba
n^{k}_{A} &=& \frac{x^k - \xi^k_{A}}{r_A} ,\ \ \
\xi^1_1 =  \frac{m_2}{m} b , \ \
\xi^1_2 = -\frac{m_1}{m} b ,\ \
\nonumber \\
\xi^2_A &=& \xi^3_A= 0.
\ea
\begin{figure}[t]
\includegraphics[scale=0.33,clip=true]{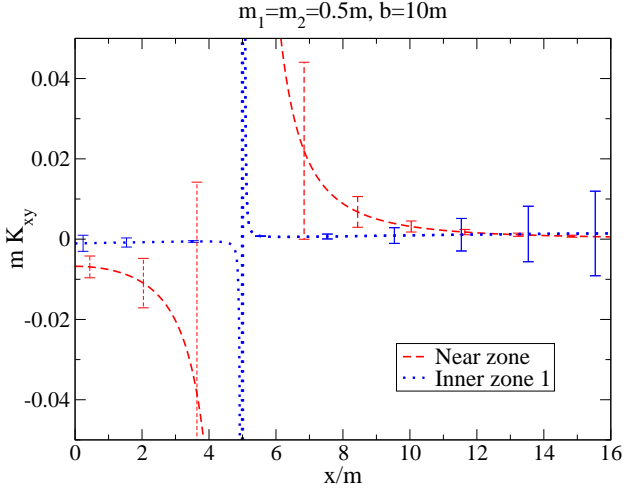}
\caption{\label{kxy} 
  This figure shows the $xy$ component of the near zone (dashed
  line) extrinsic curvature, as well as the inner zone curvature 
  (dotted line) obtained via black hole perturbation theory.
  The plot is for equal mass black holes with $m/b=1/10$.
  In the buffer zones, where the near and inner zone approximations
  have comparable error bars, the two approximations agree within 
  error bars. All other components vanish along the $x$-axis.}
\end{figure}
In Fig.~\ref{kxy} we plot the $xy$ component of the extrinsic curvature
along the positive $x$-axis in the equal mass case. For reasons of symmetry
all other components vanish along the $x$-axis. Observe that the near zone
solution (dashed line) diverges faster than the inner zone solution
(dotted line). From this figure we see that even though the resummed
ADMTT post-Newtonian expansion models well the $3$-metric of a
perturbed Schwarzschild hole, the ADMTT and inner extrinsic curvatures
do not agree well in the inner zone, just like in the case of the
shift.  However, in the buffer zone where both inner and near zone
results have similar errors, both approximations roughly agree.
On the side away from the companion black hole this agreement is within
error bars, while between the holes the two error bars do not quite overlap.
This suggests that a separation of $b=10m$ is at the edge of validity of the
matching scheme employed here. Notice, however, that the error bars
\ba
E_{KPN} &=& \frac{3}{2} v \left(\frac{m_1^2}{r_1^3} +
  \frac{m_2^2}{r_2^3} \right) + \frac{5}{4} v \frac{m_1
  m_2}{b} \left(\frac{1}{r_1^2} + \frac{1}{r_2^2} \right) 
\nonumber \\
&& - 8 v \frac{m_1 m_2}{S^3},
\nonumber \\
E_{KBHPT,A} &=& 6 \frac{m}{b} \left(\frac{r_A}{b}\right)^3
\frac{v}{r_A},
\ea
plotted here (for $v = (m/b)^{1/2}$) are only approximations 
of the true errors. These approximate errors were obtained by
differentiating the error bars for the metric in Eq.~(\ref{metric_errors})
with respect to time. Note that the absolute error in the
extrinsic curvature is $O(m/b)^{1/2}$ smaller than that of the
metric because the first non-trivial term in the extrinsic curvature
appears at $O(m/b)^{3/2}$.

\begin{figure}[t]
\includegraphics[scale=0.33,clip=true]{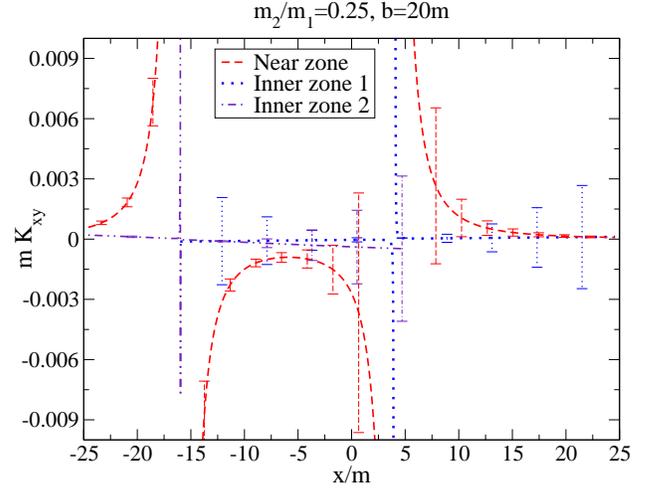}
\caption{\label{Kxy-mu} This figure shows the $xy$ component of the
  piece-wise extrinsic curvature for a system of unequal masses
  ($m_2/m_1 =0.25$) at separation $b=20m$.  Note that the scale of the
  $y$-axis has been reduced. The curves are closer to each other in
  this case because the perturbation parameter $m/b$ has decreased by
  the choice of parameters. }
\end{figure}

Similar results for the extrinsic curvature are also obtained for
different separations and mass ratios.  Figure~\ref{Kxy-mu} shows the
$xy$ component of the piece-wise extrinsic curvature along the
$x$-axis for the case of $m_2/m_1 = 0.25$ and $b=20m$.
The near zone curvature (dashed line) agrees within error bars with the
inner zone $1$ curvature (dotted line) and the inner zone $2$ curvature
(dot-dot-dashed line) in buffer zones $1$ and $2$ respectively. Observe
however, that the agreement worsens outside the buffer zone, e.g.
as we approach either hole.

\subsection*{Transition Functions}
In this subsection, we construct transition functions that allow us to
remove the piece-wise nature of Eq.~(\ref{globalpiece}) and merge the
approximate solutions smoothly in the buffer zone. This smoothing
is performed by taking weighted averages of the two approximations. 
This procedure is justified by the fact that both approximations are
equal to each other in the buffer zone up to uncontrolled
remainders (asymptotic matching theorem~\cite{Bender})
which were neglected in the approximations used.

In the following we assume that the middle of each 
buffer zone is located around
\be
\label{Wolfs_rM}
r^M_A = \left( b^4 m_A^2 / m \right)^{1/5},
\ee
which corresponds to the distance from the black holes
where the error bars of the adjacent approximations are comparable.

The transition functions we use are all of the form
\ba
\label{AttBH}
f(r) = \left\{ \begin{array}{ll}
0 ,             \qquad r \leq r_0\\
\frac{1}{2} \left\{ 1 + \tanh \left[ \frac{s}{\pi} \left( \tan(\frac{\pi}{2w}(r-r_0)) 
\right. \right. \right. \\ 
\left. \left. \left. -\frac{q^2}{\tan(\frac{\pi}{2w}(r-r_0))}\right)
\right] \right\} ,      \qquad r_0 < r < r_0+w\\
1 ,\            \qquad  r \geq r_0+w ,
\end{array} \right.
\ea
which is a function which smoothly transitions from zero to one in
the region $r_0 < r < r_0 + w $. The parameters used here are as
follows: $r_0$ defines where the transition begins; $w$ gives the
width of the transition window; $q$ determines the point where
the transition function is equal to $1/2$ (this happens at
$r_{1/2} = r_0 + (2w/\pi) \arctan q$); the slope of the transition 
function at this point is $s (1+q^2)/(2w)$ and thus can by influenced 
by choosing $s$ (and of course $w$).

\begin{figure}[t]
\includegraphics[scale=0.33,clip=true]{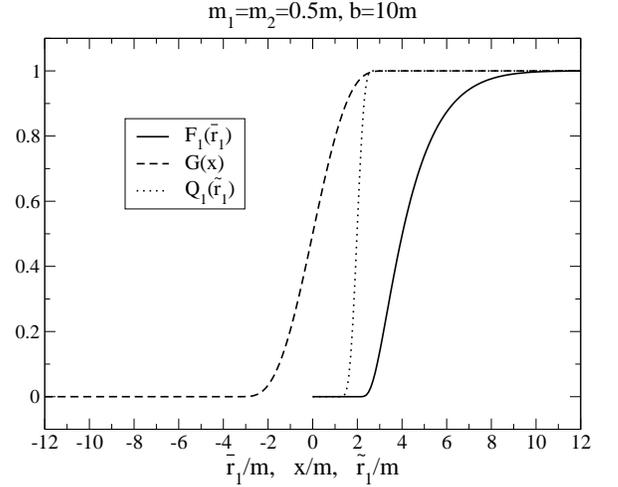}
\caption{\label{transition-func} 
  This figure shows the transition functions used to merge the
  different pieces of the $4$-metric. The transition function
  $F_1(\bar{r}_1)$ with the transition parameters of Eq.~(\ref{transFpars})
  is shown as a solid line, while the function $G(x)$ with the
  parameters of Eq.~(\ref{transFpars-2}) is shown with a dashed line.
  The transition function $Q_1(\tilde{r}_1)$ (dotted line)
  defined in Eqs.~(\ref{Q-trans}) and (\ref{Q-trans-pars})
  is used in Sec.~\ref{efcoords} to construct horizon
  penetrating coordinates via the coordinate 
  transformation~(\ref{InertialKStoCorotIso}).}
\end{figure}

The global merged 4-metric is then given by
\ba
\label{global}
g_{\mu \nu}^{(global)} &=& G(x) \left\{F_1(\bar{r}_1) g_{\mu \nu}^{(3)} +
\left[1-F_1(\bar{r}_1)\right] g_{\mu \nu}^{(1)}\right\}
\nonumber \\
&& + \left[1-G(x)\right] \left\{F_2(\bar{r}_2)g_{\mu \nu}^{(3)} +
\left[1 \right.\right.
\nonumber \\
&& \left. \left. - F_2(\bar{r}_2)\right] g_{\mu \nu}^{(2)}\right\},
\ea
where we have introduced
\be
\label{transF}
F_A(\bar{r}_A)=f(\bar{r}_A),
\ee
with
\be
\label{transFpars}
r_0 = 0.4 r^M_A, \qquad
w   = 3.5 r^M_A, \qquad
q   = 0.2, \qquad
s   = b/m ,
\ee
and also
\be
G(x)=f(x) ,
\ee
but with
\ba
\label{transFpars-2}
r_0 &=& \frac{b(m_2-m_1)}{2m} - \frac{b}{2} + 2.2 m_2 , \;
w   = b-2.2m, \; 
\nonumber \\
q   &=& 1, \;
s   = 2.5 .
\ea
The transition functions $F_1(\bar{r}_1)$ and $G(x)$ are shown
in Fig.~\ref{transition-func} for $m_A=m/2$, $b=10m$ and the
parameters of Eqs.~(\ref{transFpars}) and (\ref{transFpars-2}),
respectively.

The global lapse, shift and extrinsic curvature are merged using the
same transition functions as for the 4-metric above.  With these
transition functions, the global metric is a mixture of inner zone and
near zone metric in a transition region given by $0.4 r^M_A < \bar{r}_A <
3.9 r^M_A$. On the other hand, the global metric becomes identical to
the inner zone metric in the region $\bar{r}_A < 0.4 r^M_A$, while it
becomes equal to the near zone metric in the region $\bar{r}_A > 3.9 r^M_A$.
Note that although the transition function are identical to those used
in paper $1$, the transition parameters chosen here are slightly
different. This change is because the inner zone metric is very
similar to the near zone metric close to the black holes in ADMTT
coordinates, so that the transition region has been moved closer to
the black holes.

The exact choice of a transition function is to a certain degree
arbitrary and could in principle be changed. However, the resulting
global metrics generated by any other reasonable transition function
should look very similar, and in fact be identical up to uncontrolled
higher order post-Newtonian and tidal perturbation terms. Also note
that the functions presented in this section are general enough to
perform a smooth merge for systems with different masses and
separations.

\begin{figure}[t]
\includegraphics[scale=0.33,clip=true]{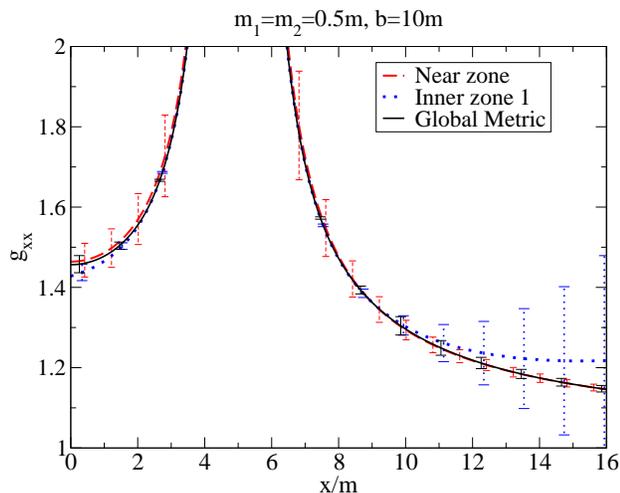}
\caption{\label{transitionfuncxx} 
  In this figure we show how the transition function takes the $xx$
  component of the near zone metric to the inner zone solution around
  BH~1. Observe that the transition function is smooth and does not
  introduce kinks into the global solution.}
\end{figure}
In Fig.~(\ref{transitionfuncxx}), we present the ${xx}$-component of
the metric around BH~1. In this figure, the dotted curve corresponds
to the inner zone metric, the dashed line is the near zone metric and
the solid line is the merged global metric. The error bar in the
global metric is given by the smallest of the error bars of the
respective approximations. The purpose of this figure is to show
explicitly that the transition function effectively merges the
different approximations in the buffer zone, where the errors are
comparable.

\begin{figure}[t]
\includegraphics[scale=0.33,clip=true]{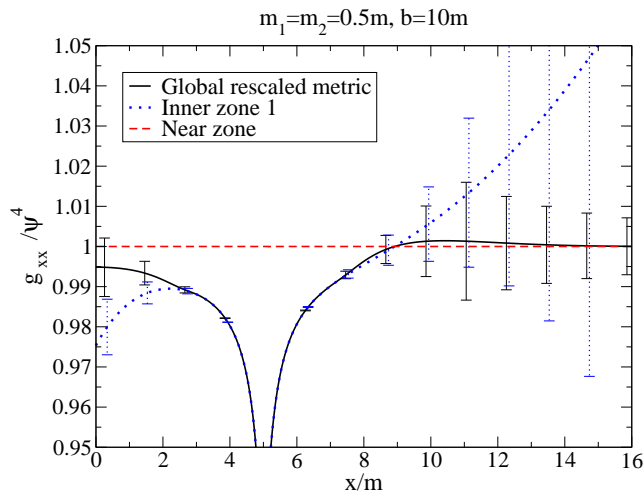}
\caption{\label{gxx-psi} 
  In this figure we show the x-component of the metric, divided by
  $\Psi^4$. The  dashed line is the near zone metric, the 
  dotted line is the inner zone metric and the black solid line is the
  global metric with transition functions.} 
\end{figure}

Even though the near zone 3-metric models the inner zone
3-metric well near the black hole horizons,
the two metrics diverge differently at $r_A=0$.  In
Fig.~\ref{gxx-psi}, we plot the xx-component of the metric divided by
$\Psi^4$. This denominator removes the divergence of the near zone
metric, which now becomes identically unity, while showing that
the inner zone metric differs by an amount equal to the size of
the tidal perturbation near the horizon.
The reason that the inner zone metric still diverges after dividing by
$\Psi^4$ is due to the fact that the tidal perturbation
piece of the inner zone metric is divergent at $r_A=0$.
This happens because the tidal perturbation
piece is valid only for finite space-like separations from the horizon,
while $r_A=0$ which is located at the inner asymptotically flat
end inside the black hole, is infinitely far from the horizon.
This means that the tidal perturbation piece of the inner zone metric is not
valid near $r_A=0$. Yet, if this spurious tidal perturbation is removed, the
resulting $3$-metric of the Schwarzschild background will again agree well
with the resummed post-Newtonian near zone $3$-metric.

\begin{figure}[t]
\includegraphics[scale=0.33,clip=true]{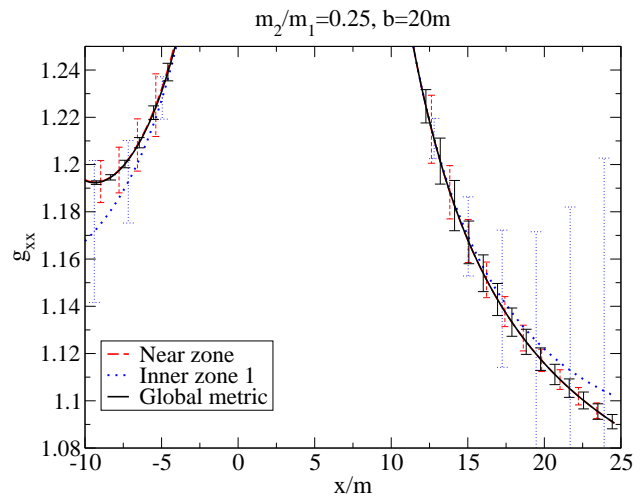}
\caption{\label{transitionfuncxx-mu} 
In this figure we plot a system of unequal masses ($m_2/m_1 = 0.25$)
and a larger separation ($b = 20 m$). We use the same parameters in
the transition function as in the equal mass case.}
\end{figure}

The same analysis can be performed for systems of unequal masses and at
different separations. In Fig.~\ref{transitionfuncxx-mu} we plot
the $xx$ component of the metric along the $x$-axis for a binary with
mass ratio $m_2/m_1 = 0.25$ and at a separation $b = 20 m$.  
The transition is now easier than before (note the scale of the
$y$-axis), because the perturbation parameter has become
smaller. Also note that we have used the same parameters in the
transition function as those used in the equal mass case and the
transition is still smooth.

\begin{figure}
\includegraphics[scale=1.0,clip=true]{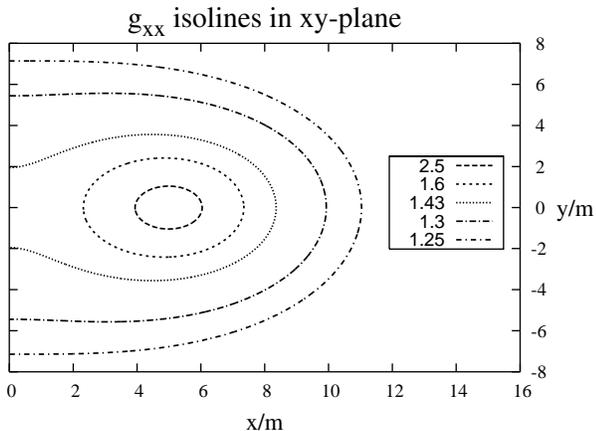}
\caption{\label{globalxx-cont}
  Contour plot of the $xx$-component of the global metric in the
  $xy$-plane around black hole $1$ for the equal mass system. Apart
  from the singularity at $x=\pm 5m$, $y=0$ the 3-metric is everywhere
  smooth. The different line styles correspond to different isolines
  of constant metric value.}
\end{figure}

\begin{figure}
\includegraphics[scale=1.0,clip=true]{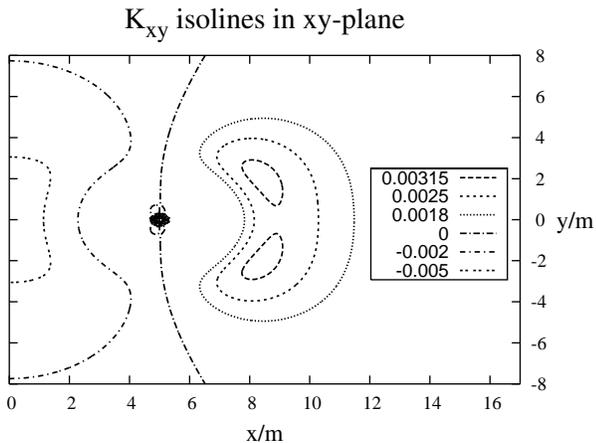}
\caption{\label{kxyglobal-cont}
  Contour plot of the $xy$ component of the global extrinsic curvature
  in the $xy$-plane around hole $1$ for the equal mass system. The
  different line styles correspond to different isolines of constant
  extrinsic curvature.}
\end{figure}

With these transition functions we can construct a smooth global
metric, as shown in Fig.~\ref{globalxx-cont}. There are no
discontinuous features in the global metric.
Just as in the case of the metric, to generate a global extrinsic
curvature we will use the same transition function of the previous
section with the same parameters. Figure~\ref{kxyglobal-cont} shows the
global extrinsic curvature with the transition functions.
Note that we could have computed the extrinsic curvature directly from
Eq.~(\ref{global}), but this curvature would have been very similar to
the curvature merged with transition functions. The difference between
these two curvatures lies in the derivatives of the transition
functions, but these terms will be of the same order or smaller than
the uncontrolled remainders in either approximation in the buffer zone
because of the parameter choice in Eqs.~(\ref{transFpars})
and~(\ref{transFpars-2}).

The 3-metric $g_{ij}^{(global)}$ and the extrinsic curvature
given in this section can now
be used as initial data for black hole binaries. Recall, however, that
these data are solutions to the Einstein equations only approximately
and, thus, they only approximately satisfy the constraints. An
analytic estimate of the constraint violation and physical accuracy of
this data is given in Table II.
\begin{table}[htb]
\begin{center}
\medskip
\begin{tabular}{||l|l||l||}
\hline
\hline
Zone & Constraint Viol. & Physical Error \\
\hline\hline
Inner zone BH$1$ (${\cal{C}}_1$) & $O[(m/b)^{5/2}(\bar{r}_1/b)^2]$ & $O[(m/b)(\bar{r}_1/b)^3]$ \\
Inner zone BH$2$ (${\cal{C}}_2$) & $O[(m/b)^{5/2}(\bar{r}_2/b)^2]$ & $O[(m/b)(\bar{r}_2/b)^3]$ \\
Near zone (${\cal{C}}_3$) & $O(m/r_A)^2$ &  $O(m/r_A)^2$ \\
\hline\hline
\end{tabular}
\\[10pt]
{Table II: This table shows an order of magnitude estimate of the
  constraint violation of this data, together with the error in the
  physics.}
\end{center}

\end{table}
Note that in the near zone, both the errors in the physics and the
constraint violations are given by the neglected terms in the
post-Newtonian approximation, which scale as $O(m/r_A)^2$. On the other
hand, in the inner zone, the leading error in the physics is due to
the terms neglected in the perturbation, which scales as
$O[(m/b)(r/b)^3]$. The inner zone constraint violations, however, are
smaller, because the perturbed Schwarzschild metric used here, 
satisfies the Einstein Equations up to order $O[(m/b)^{5/2}(r/b)^2]$.

In order to obtain initial data that satisfy the constraints exactly,
it will be necessary to project the data given in this paper to the
constraint hypersurface. However, since these data are already
significantly close to this hypersurface, sensible projection methods
should {\textit{not}} alter much the astrophysical content of the
initial data. Furthermore, as stressed earlier, if this constraint
violation is smaller than discretization error, these data could be
evolved without any projection.

There are a couple of caveats that need to be discussed in more
detail. First, as in paper $1$, asymptotic matching will break down
when the separation of the bodies is small enough that the near zone
disappears between the two black holes.  An approximate criterion for
when this happens, corresponds to the separation where the middles of
the two buffer zone shells touch for the first time.  If we define the
middle of each buffer zone as in Eq.~(\ref{Wolfs_rM}), this happens
when the two spheres of radius $r^M_1$ and $r^M_2$ centered on each of
the black holes, touch for the first time 
(roughly $b \approx 8m$ for an equal mass system).
Furthermore, the tidal perturbation used in the inner
zone metric is only valid for small spacelike separation from the
horizon. This criterion implies that the tidal perturbation is good
roughly for
\begin{equation}
{m_A \over 2} {m \over 2b} \ll {\bar{r}}_A \ll {m_A \over m} b.
\end{equation}
Therefore, numerical simulations will need to excise the data
somewhere inside the apparent horizons of the black holes before
evolving it, since the inner zone metric is not valid all the way up
to $\bar{r}_A =0$.  Notice, however, that these limitations are due to
the approximations and coordinates used and not due to asymptotic
matching.

\section{Constraint Violations} 
\label{constraints}

The initial data constructed here are based on approximate solutions
and, thus, they do not solve the Einstein equations exactly. We have
estimated that the largest error in the constraints of the full theory
occurs in the buffer zone and is at most $O(m_A/b)^2$.  This error can
be sufficiently small compared to other sources of numerical error
such that solving the constraints more accurately may not be required.
On the other hand, the initial data could be used as input in one of
the many conformal decompositions~\cite{Cook:2000vr} to compute a
numerical solution to the full constraints.
In this section, we study the constraint violations of the data
presented in this paper and we compare them to what was obtained in
paper $1$. 

\begin{figure}[t]
\includegraphics[scale=0.33,clip=true]{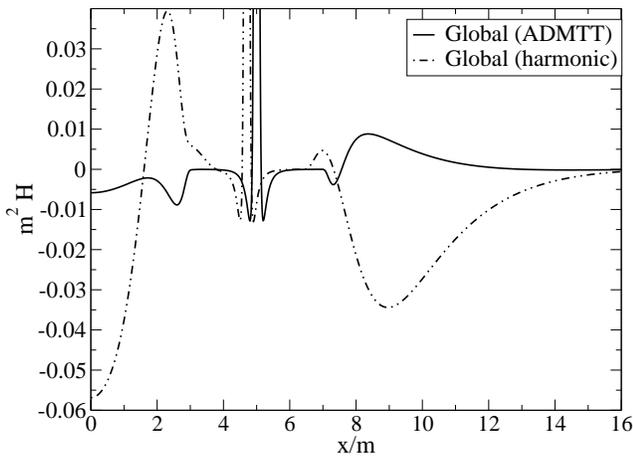}
\caption{\label{h-cons} 
  This figure shows the violation of the Hamiltonian constraint for
  the equal mass system along the positive $x$-axis for both the data
  presented in this paper (solid black line) and that of paper $1$
  (dot-dashed black line).}
\end{figure}
In Fig.~\ref{h-cons} we show the Hamiltonian constraint violation
along the positive $x$-axis for both data sets. The solid line is the
constraint violation produced by the data presented in this paper,
while the dot-dashed line is that produced by the data of paper $1$.
Observe that far away from both holes, both data sets (paper $1$ and
this paper's) have a constraint violation that approaches zero. This
feature arises because both global metrics are given by a $1$ PN
solution in the near zone, which has a constraint violation that
decays when the expansion parameter $m_A/r_A$ becomes smaller.
Note, however, that the constraint violation in
ADMTT gauge is much smaller than in harmonic gauge. The reason for
this difference is that our resummed PN expressions satisfy
the Hamiltonian constraint up to errors of $O(m_A/r_A)^3$ in ADMTT
gauge, while the harmonic gauge expressions have errors 
of $O(m_A/r_A)^2$.

Also in Fig.~\ref{h-cons}, observe that close to the holes, the
violation is given by that of the inner zone solution, which is small
near the horizon, but increases as we move away from the holes. The
largest constraint violation occurs in the buffer zone, where we
transition from inner to near zone expressions. Note that this maximum
in the violation is mainly caused by the transition functions.  
\begin{figure}[t]
\includegraphics[scale=0.33,clip=true]{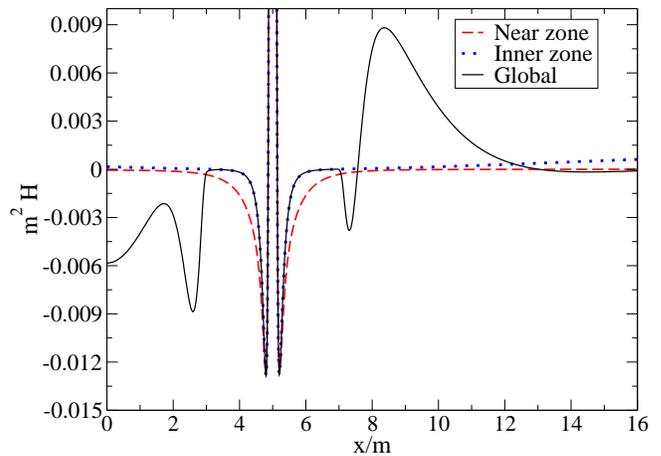}
\caption{\label{ham-inner-near-global} This figure shows the violation
  of the Hamiltonian constraint for the equal mass system along the
  positive $x$-axis for the near zone (dashed line), the inner zone
  (dotted line) and the global (solid line) data with transition
  functions.}
\end{figure}
As can be seen in Fig.~\ref{ham-inner-near-global}, both the inner and near
zone solutions individually have smaller constraint violations in the buffer
zone than the global curve.
This, however, is not an indication of a bad choice of transition
functions. Rather the inner and near zone solutions have smaller
constraint violations than expected, because we have kept some extra
higher order terms. Recall that matching in the buffer zone was only
performed up errors of $O(m_A/r_A)^{2}$. Hence the two solutions
differ by this amount, and thus the error in the physics is of the
same order as well. Nevertheless, the inner zone solution has
constraint violations of $O(m_A/r_A)^{5/2}$ only, and the errors in
the Hamiltonian constraint in the ADMTT near zone solution are only of
$O(m_A/r_A)^3$. When these two solutions (which are equal to up to
order $O(m_A/r_A)$) are averaged with a transition function we obtain
a new solution which differs from both inner and near zone solution by
$O(m_A/r_A)^{2}$. Therefore, we expect the error in the constraints
of the averaged solution to be of $O(m_A/r_A)^{2}$ as well, and thus
larger than the error in the individual solutions.

\begin{figure}
\includegraphics[scale=0.33,clip=true]{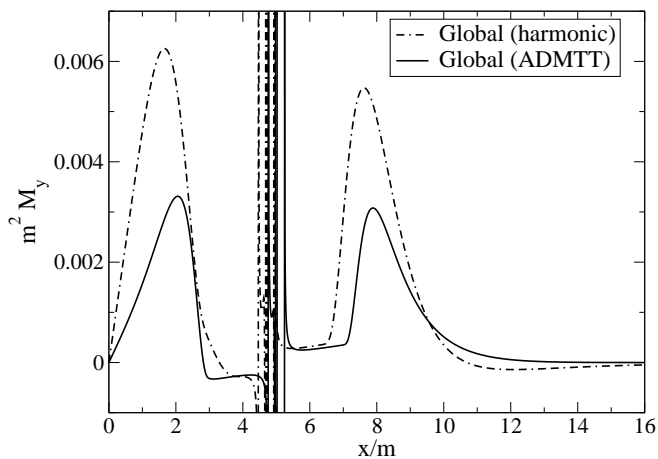}
\caption{\label{p-cons} 
  This figure shows the $y$-component of the momentum constraint
  violation along the positive $x$-axis for the data computed in this
  paper (solid black line) and in paper $1$ (dot-dashed black line),
  assuming the parameters of the equal mass system. Note that all
  other components vanish along this axis.}
\end{figure}

From Fig.~\ref{h-cons} we see that the Hamiltonian
constraint for the data of this paper is smaller than the
data of paper $1$. This decrease in the constraint violation is an
indication that the matching performed in this paper produces near and inner
zone solutions that are closer to each other in the buffer zone. Thus, the
transition function has to do less work to join the solutions together,
therefore introducing less of a constraint violation. We should point out
that while the functional form of the transition function in paper $1$ and
in this paper is identical, the parameters used are different. The biggest
difference is that here we use a smaller transition window $w$. Thus
stronger artificial gradients and larger constraint violations might be
expected in this work. This however, does not happen because matching works
so much better in ADMTT coordinates that the inner and near zone solutions
are substantially closer.
Also note that the singularities in the constraint violations of the
data in ADMTT and harmonic gauge (see Figs.~\ref{h-cons} and
\ref{p-cons}) occur at different coordinate locations.
This is simply because the inner zone metric, which
is relevant in this region, is expressed in different coordinates.

One obtains qualitatively similar plots if the constraints are plotted
along different directions.
\begin{figure}
\includegraphics[scale=1.0,clip=true]{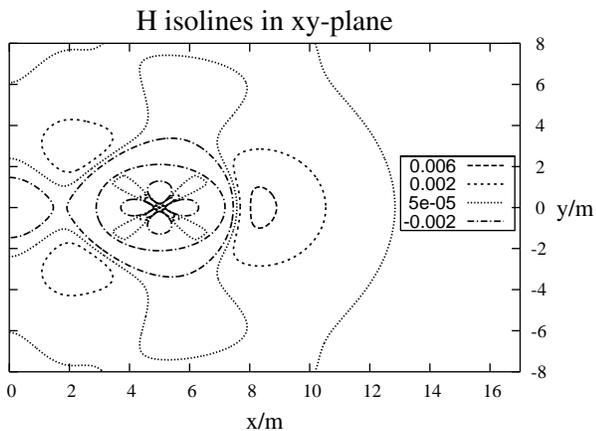}
\caption{\label{ham-cont} Hamiltonian constraint
  violation of the global metric in ADMTT coordinates in the
  $xy$-plane for the equal mass system. The different line styles
  corresponds to isolines of constant constraint violation.}
\end{figure}
Evidence for this behavior can be seen in Fig.~\ref{ham-cont}, which shows
a contour plot of Hamiltonian constraint violation for the global metric in
ADMTT coordinates in the $xy$-plane. We see that there is ring of radius
$3m$ of negative constraint violation around the black hole. Outside this
ring, about $4.5m$ away from the hole, there are three maxima. The largest
of these occurs on the $x$-axis. Note that these minima and maxima are all
located in the buffer zone. In addition, there is a blow-up inside the
horizon at $r_A=0$.

In Fig.~\ref{p-cons}, we plot the $y$-component of the momentum
constraint along the positive $x$-axis close to black hole $1$ for both
data sets. For reasons of symmetry all other components vanish
along this axis.
Observe that the violation is everywhere small, reaching a maximum in the
buffer zone. As expected, in the near zone the
violation decays as we move away from the black holes and becomes
identically post-Newtonian. Note however, that the resummed 
PN data in ADMTT gauge satisfy the momentum constraint exactly,
while in harmonic gauge it has an error of $O(m_A/r_A)^2$.
On the other hand, in the inner zone, the violation
is close to zero outside the horizon (which is located at $r_A \approx
m_A/2$) and it grows as $r_A$ becomes larger. As in the case of the
Hamiltonian constraint the largest violation occurs in the region where the
transition function leads to non-trivial averaging of the two
approximations. The violation is once more smaller for the data 
presented in this paper relative to that of paper $1$, 
which is one more indication that the matching is smoother here.
Finally, observe that close to the holes, and in particular close to the
horizons, the violation due to the inner zone data diverges. This divergence
can be traced back to the choice of slicing, which forces the lapse to be
zero at the horizon.

\begin{figure}
\includegraphics[scale=0.33,clip=true]{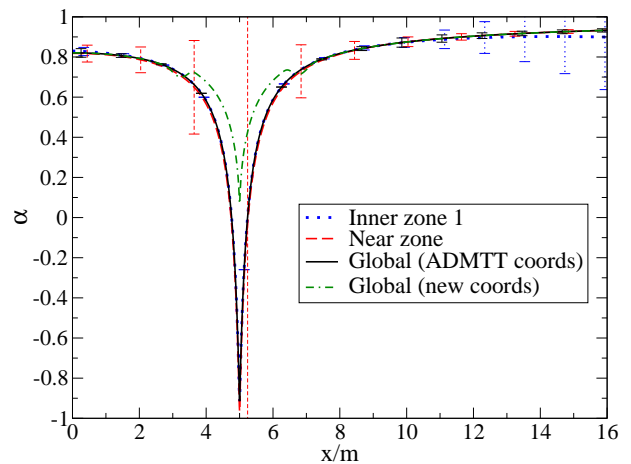}
\caption{\label{lapse-collapse} This figure shows the near zone (dashed
  line), the inner zone (dotted line) and the global lapse (solid
  line) along the positive x-axis in ADMTT coordinates for the equal
  mass system. Observe that the global ADMTT lapse crosses zero near
  the horizon. Also shown is the lapse (dash-dash-dotted line) in the
  new horizon-penetrating coordinates of Sec.~\ref{efcoords}.}
\end{figure}
The vanishing of the lapse can be observed in
Fig.~\ref{lapse-collapse}, where we plot the inner zone, near zone and
global lapse along the positive $x$-axis. This figure shows how the
inner zone lapse goes through zero near the horizon. The vanishing of
the lapse translates into a divergence of the momentum constraint,
which renders excision difficult and the data difficult to implement
numerically.  However, this is a failure of the coordinate system used
in the inner zone and not of the method of asymptotic matching. In the
next section, we construct a new coordinate system in which the lapse
remains non-zero across the horizon, and approaches an isotropic
Schwarzschild lapse in the buffer zone to allow for matching.

In Figs.~\ref{h-cons} and \ref{p-cons} we have compared the constraint
violation of the data set presented here to that presented in paper
$1$. Technically this comparison suffers from the problem that the
constraints were computed in different coordinate systems.  Note
however, that both coordinate systems are identical to leading order.
Yet instead of comparing the violations point by point, we can study
their global average. Figs.~\ref{h-cons} and \ref{p-cons} favor the
data of this paper because the humps are consistently smaller. As
explained earlier, this is an indication that after performing the
matching, the inner and near zone solutions are very close to each
other, making the transition smoother.

\section{Horizon penetrating coordinates}
\label{efcoords}

Recall that the global coordinate system constructed in 
Sec.~\ref{matching} was based on matching the tidally perturbed black hole
metric of the inner zone in isotropic coordinates to a post-Newtonian metric
in ADMTT coordinates in the near zone. We have chosen to match in these
coordinates because they are already very similar to each other.
Hence, the coordinate transformation [Eq.~(\ref{fulltransf-nokappa})] that
leads to matching is close to the identity, and thus facilitates
computations. However, the fact that the
global coordinate system remains very close to isotropic coordinates is also
a disadvantage, since the $t=0$ slice is not horizon penetrating in
isotropic coordinates. The lapse goes through zero very close to the
horizon and in addition, the extrinsic curvature has a coordinate singularity
at the point where the lapse goes through zero. Initial data on such a
slice may not be suitable for numerical evolutions. In this section we
present a remedy for this problem. By constructing a further coordinate
transformation, we obtain coordinates which are close to horizon
penetrating Kerr-Schild coordinates near each black hole.

The basic strategy we will use is the following. We first
determine the perturbed black hole metric valid in the inner zone
in Kerr-Schild coordinates. 
We then transform this metric into a coordinate system
which remains Kerr-Schild near the black hole horizons, but which is
corotating isotropic in the buffer zone. In this manner, the
transformed inner zone metric components in the buffer zone will
remain identical to those in Sec.~\ref{matching}, so that we can use
the same matching coordinate transformation
[Eq.~(\ref{fulltransf-nokappa})] as in Sec.~\ref{matching}.

The standard transformation between spherical Kerr-Schild
$(\hat{t}',\hat{r}_1')$ and spherical isotropic coordinates
$(\bar{t},\bar{r}_1)$ centered at hole $1$ and its inverse are given by
\cite{MTW}
\ba
\label{ks-map}
\hat{t'} &=& \bar{t} + 2 M_1 \ln{\left[\frac{\bar{r}_1}{2 M_1}
 \psi(\bar{r}_1)^2 -1 \right]}, \qquad  \hat{r}_1' = \bar{r}_1 \psi(\bar{r}_1)^2,
\nonumber \\
\bar{t} &=& \hat{t}' - 2 M_1 \ln{\left[ \frac{\hat{r}_1'}{2 M_1} -1 \right]},
\qquad \bar{r}_1 =  \frac{\hat{r}_1'}{\psi(\hat{r}_1')^2},
\nonumber \\
\bar{\theta} &=&\hat{\theta}'  , \qquad  \bar{\phi} = \hat{\phi}', 
\ea
where the radius is measured from the center of a black hole with mass
$M_1$ and where the conformal factor and its inverse are given by 
\ba
\label{Brill-Lindq}
\psi(\bar{r}_1) &=& 1 + \frac{M_1}{2 \bar{r}_1},
\nonumber \\
\psi(\hat{r}_1') &=& \frac{\hat{r}_1'}{M_1} \left[ 1 - \left( 1 -
  \frac{2M_1}{\hat{r}_1'}\right)^{1/2} \right].
\ea
Note that the inverse transformation contains a square root in the
conformal factor, which becomes complex inside the
horizon. For this reason, it is simpler to first transform the inner
zone metric to Kerr-Schild coordinates analytically. Then, we can
construct a new coordinate system that is the identity map near the
horizon, but that brings back the metric to isotropic coordinates
outside the horizon in the buffer zone.

Let us first transform the inner zone metric of hole $1$ to
Kerr-Schild coordinates. By applying Eqs.~(\ref{ks-map}) to the inner
zone metric in spherical isotropic coordinates [Eq.~($3.22$) of
Ref.~\cite{Alvi:1999cw}] we obtain
\begin{widetext}
\ba
\label{gab-ks-sph}
g_{\hat{0}' \hat{0}'}^{(1)} &=& -f + \frac{m_2}{b^3} \hat{r}_1'^2
f^2 \left[ 3 \sin^2{\hat{\theta}'} \left( \cos{\hat{\phi}'}
  \cos{\Omega \bar{t}} + \sin{\hat{\phi}'} \sin{\Omega \bar{t}}
  \right)^2 - 1 \right],
\nonumber \\
g_{\hat{1}' \hat{1}'}^{(1)} &=& \left(2-f\right) + \frac{m_2}{b^3} \hat{r}_1'^2
\left(1 + \frac{4 M_1^2}{\hat{r}_1'^2} \right)  \left[ 3 \sin^2{\hat{\theta}'}
  \left( \cos{\hat{\phi}'} \cos{\Omega \bar{t}} + \sin{\hat{\phi}'}
  \sin{\Omega \bar{t}} \right)^2 - 1 \right],
\nonumber \\
g_{\hat{2}' \hat{2}'}^{(1)} &=& \hat{r}_1'^2 + \frac{m_2}{b^3} \hat{r}_1'^2
f^2 \left[ 3 \sin^2{\hat{\theta}'}  \left( \cos{\hat{\phi}'}
  \cos{\Omega \bar{t}} + \sin{\hat{\phi}'} 
  \sin{\Omega \bar{t}} \right)^2 - 1 \right],
\nonumber \\
g_{\hat{3}' \hat{3}'}^{(1)} &=& g_{\hat{2}' \hat{2}'}^{(1)}
\sin{\hat{\theta}'},
\nonumber \\
g_{\hat{0}' \hat{1}'}^{(1)} &=& \frac{2 M_1}{\hat{r}_1'} -
\frac{m_2}{b^3} \hat{r}_1'^2 f \frac{2 M_1}{\hat{r}_1'}  \left[ 3
  \sin^2{\hat{\theta}'}  \left( \cos{\hat{\phi}'} \cos{\Omega \bar{t}}
  + \sin{\hat{\phi}'} \sin{\Omega \bar{t}} \right)^2 - 1 \right],
\nonumber \\
g_{\hat{0}' \hat{2}'}^{(1)} &=& - \frac{2 m_2}{b^3} \sqrt{\frac{m}{b}}
\hat{r}_1'^3 f \cos{\hat{\theta}'} \left( \sin{\hat{\phi}'}
\cos{\Omega \bar{t}} - \cos{\hat{\phi}'} \sin{\Omega \bar{t}} \right),
\nonumber \\
g_{\hat{0}' \hat{3}'}^{(1)} &=& - \frac{2 m_2}{b^3} \sqrt{\frac{m}{b}}
\hat{r}_1'^3 f \sin{\hat{\theta}'} \cos{2 \hat{\theta}'} \left(
\cos{\hat{\phi}'} \cos{\Omega \bar{t}} + \sin{\hat{\phi}'} \sin{\Omega
  \bar{t}} \right), 
\nonumber \\
g_{\hat{1}' \hat{2}'}^{(1)} &=& \frac{1}{f} g_{\hat{0}' \hat{2}'}^{(1)},
\qquad g_{\hat{1}' \hat{3}'}^{(1)} = \frac{1}{f} g_{\hat{0}'
  \hat{3}'}^{(1)},
\ea
\end{widetext}
where all coordinates are centered on BH~1 and where we used the
abbreviation
\be
f = \left(1 - \frac{2 M_1}{\hat{r}_1'}\right).
\ee
In Eq.~(\ref{gab-ks-sph}), $\bar{t}$ stands for the isotropic time
coordinate, given in Eq.~(\ref{ks-map}).
Recall that this metric was derived under the assumption that
the second black hole responsible for the tidal perturbation
is moving slowly. In particular Alvi obtained the perturbation
from a stationary perturbation by replacing $\bar{\phi}$ by
$\bar{\phi} - \Omega \bar{t}$. This means the largest error in this
perturbation is of order $O(m/b)^{5/2} O(r_A/b)^2$.
In the following we will replace Schwarzschild time $\bar{t}$
by Kerr-Schild time $\hat{t}$ to simplify our expressions.
This replacement will change the tidal perturbation only at order
$O(m/b)^{5/2} O(r_A/b)^2$ and thus not introduce any extra errors.

We now go one step further and transform Eq.~(\ref{gab-ks-sph}) to
Cartesian Kerr-Schild coordinates via the standard map 
\ba
\hat{t}' &=& \hat{t}, \qquad \hat{r}_1' = \left(\hat{x}^2 + \hat{y}^2
+\hat{z}^2 \right)^{1/2},
\nonumber \\
\hat{\theta}' &=& \cos^{-1}{\left[\frac{\hat{z}}{\left(\hat{x}^2 + \hat{y}^2
+\hat{z}^2 \right)^{1/2}}\right]},
\nonumber \\
\hat{\phi}' &=& \tan^{-1}{\left(\frac{\hat{y}}{\hat{x}}\right)}.
\ea
We then obtain
\begin{widetext}
\ba
\label{gab-ks-car}
g_{\hat{0} \hat{0}}^{(1)} - g_{\hat{0} \hat{0}}^{(KS)} &=& \frac{m_2}{b^3} f^2 d,
\nonumber \\
g_{\hat{1} \hat{1}}^{(1)} - g_{\hat{1} \hat{1}}^{(KS)} &=& \frac{m_2}{b^3} d
\left[1 + \frac{2 M_1^2}{\hat{r}_1^2} \left(\frac{3
      \hat{x}^2}{\hat{r}_1^2} -1\right)\right] -  \frac{4 M_1}{\hat{r}_1}
\frac{m_2}{b^3}  \sqrt{\frac{m}{b}} \frac{\hat{x}}{\hat{r}_1}
\left[\left( \hat{z}^2 - \hat{y}^2\right) \sin{\Omega \hat{t}} -
  \hat{y} \hat{x} \cos{\Omega \hat{t}} \right], 
\nonumber \\
g_{\hat{2} \hat{2}}^{(1)} - g_{\hat{2} \hat{2}}^{(KS)} &=& \frac{m_2}{b^3} d
 \left[1 + \frac{2 M_1^2}{\hat{r}_1^2} \left(\frac{3 \hat{y}^2}{\hat{r}_1^2} -1\right)\right] -
 \frac{4 M_1}{\hat{r}_1} \frac{m_2}{b^3} 
 \sqrt{\frac{m}{b}} \frac{\hat{y}}{\hat{r}_1} \left[\left( \hat{x}^2 -
  \hat{z}^2\right) \cos{\Omega \hat{t}} + \hat{y} \hat{x} \sin{\Omega
    \hat{t}} \right], 
\nonumber \\
g_{\hat{3} \hat{3}}^{(1)} - g_{\hat{3} \hat{3}}^{(KS)} &=& \frac{m_2}{b^3} d
 \left[1 + \frac{2 M_1^2}{\hat{r}_1^2}
   \left(\frac{3 \hat{z}^2}{\hat{r}_1^2} -1\right)\right] -
 \frac{4 M_1}{\hat{r}_1} \frac{m_2}{b^3} 
 \sqrt{\frac{m}{b}} \frac{\hat{z}^2}{\hat{r}_1} \left( \hat{y}
\cos{\Omega \hat{t}} - \hat{x} \sin{\Omega \hat{t}} \right), 
\nonumber \\
g_{\hat{0} \hat{1}}^{(1)} - g_{\hat{0} \hat{1}}^{(KS)} &=& - \frac{2 M_1}{\hat{r}_1}
\frac{\hat{x}}{\hat{r}_1} \frac{m_2}{b^3} f d + \frac{2 m_2}{b^3}
\sqrt{\frac{m}{b}} f \left[ \left(\hat{z}^2 - \hat{y}^2\right)
  \sin{\Omega \hat{t}}  - \hat{x}\hat{y} \cos{\Omega \hat{t}}\right], 
\nonumber \\
g_{\hat{0} \hat{2}}^{(1)} -  g_{\hat{0} \hat{2}}^{(KS)}&=& - \frac{2 M_1}{\hat{r}_1}
\frac{\hat{y}}{\hat{r}_1} \frac{m_2}{b^3} f d + \frac{2 m_2}{b^3}
\sqrt{\frac{m}{b}} f \left[ \left(\hat{x}^2 - \hat{z}^2\right)
  \cos{\Omega \hat{t}} + \hat{x}\hat{y} \sin{\Omega \hat{t}} \right],
\nonumber \\
g_{\hat{0} \hat{3}}^{(1)} - g_{\hat{0} \hat{3}}^{(KS)} &=& - \frac{2 M_1}{\hat{r}_1}
\frac{\hat{z}}{\hat{r}_1} \frac{m_2}{b^3} f d + \frac{2 m_2}{b^3}
\sqrt{\frac{m}{b}} f \hat{z} \left( \hat{y} \cos{\Omega \hat{t}} -
\hat{x} \sin{\Omega \hat{t}} \right),
\nonumber \\
g_{\hat{1} \hat{2}}^{(1)} - g_{\hat{1} \hat{2}}^{(KS)}  &=&
\frac{\hat{x}\hat{y}}{\hat{r}_1^2} \frac{6 m_2}{b^3} \frac{M_1}{\hat{r}_1^2} d 
  + \frac{4 m_2}{b^3} \frac{M_1}{\hat{r}_1} \sqrt{\frac{m}{b}} 
  \frac{1}{\hat{r}_1^3} \left[\left(\hat{y} \sin{\Omega \hat{t}} +
      \hat{x} \cos{\Omega \hat{t}} \right)
    \left(\hat{y}^4 - \hat{x}^4 \right) 
+ \left(\hat{x} \cos{\Omega \hat{t}}  - \hat{y} \sin{\Omega \hat{t}}
    \right) \hat{z}^2 \left( 2 \hat{y}^2 + \hat{z}^2 \right)\right],  
\nonumber \\
g_{\hat{1} \hat{3}}^{(1)} - g_{\hat{1} \hat{3}}^{(KS)} &=&
\frac{\hat{x}\hat{z}}{\hat{r}_1^2} \frac{6 m_2}{b^3} \frac{M_1^2}{\hat{r}_1^2} d
 + \frac{4 m_2 M_1}{b^3} \sqrt{\frac{m}{b}} \left(1 - \frac{2
  \hat{z}^2}{\hat{r}_1^2} \right) \hat{z} \sin{\Omega \hat{t}},
\nonumber \\
g_{\hat{2} \hat{3}}^{(1)} - g_{\hat{2} \hat{3}}^{(KS)} &=&
\frac{\hat{y}\hat{z}}{\hat{r}_1^2} \frac{6 m_2}{b^3} \frac{M_1^2}{\hat{r}_1^2} d
 - \frac{4 m_2 M_1}{b^3} \sqrt{\frac{m}{b}} \left(1 - \frac{2
  \hat{z}^2}{\hat{r}_1^2} \right) \hat{z} \cos{\Omega \hat{t}},
\ea
\end{widetext}
where $\hat{r}_1 = \hat{r}_1'$.
This is the inner zone metric for hole $1$ in Cartesian Kerr-Schild
coordinates, where the standard metric in Cartesian Kerr-Schild form
is
\be
g_{\hat{a}\hat{b}}^{(KS)} = \eta_{\hat{a} \hat{b}}  + \frac{2
  M_1}{\hat{r}_1} l_{\hat{a}} \l_{\hat{b}}, 
\ee
with null vectors $l_{\hat{a}} = \{1,x^{\hat{i}}/\hat{r}_1\}$ and where
we have introduced the shorthand 
\be
d = \left[ 3 \left(\hat{x} \cos{\Omega \hat{t}} + \hat{y}
  \sin{\Omega \hat{t}} \right)^2 - \hat{r}_1^2 \right].
\ee
The lapse in this coordinate system is now a positive definite
function.

The matching of Sec.~\ref{matching}, however, is performed in the
buffer zone in Cartesian corotating isotropic coordinates. We thus
need a coordinate transformation that leaves Eq.~(\ref{gab-ks-car})
unchanged near the horizon, but takes the metric to isotropic
corotating coordinates in the buffer zone. 
For black hole A, the transformation we use is given by
\ba
\label{InertialKStoCorotIso}
\hat{t} &=& \left(1-Q_A\right) \tilde{t} + Q_A \left[
  \tilde{t} + 2 m_A \ln{\left|\frac{\tilde{r}_A\psi_A^2}{2 m_A} -1 \right|} \right],
\nonumber \\
\hat{x} &=& \left(1-Q_A\right) \tilde{x} + Q_A \;
\left(\tilde{x} \cos\Omega\tilde{t} - \tilde{y} \sin\Omega\tilde{t} \right)\;
\psi_A^{2},
\nonumber \\
\hat{y} &=& \left(1-Q_A\right) \tilde{y} + Q_A
\;\left(\tilde{x} \sin\Omega\tilde{t} + \tilde{y} \cos\Omega\tilde{t}\right)\;
\psi_A^{2},
\nonumber \\
\hat{z} &=& \left(1-Q_A\right) \tilde{z} + Q_A
\;\tilde{z}\; \psi_A^{2},
\ea
where 
\be
\psi_A = 1 + \frac{m_A}{2 \tilde{r}_A}, \qquad
\tilde{r}_A = \sqrt{\tilde{x}^2+\tilde{y}^2+\tilde{z}^2}
\ee
and the transition function
\be
\label{Q-trans}
Q_A = f(\tilde{r}_A)
\ee
with
\ba
\label{Q-trans-pars}
r_0 &=& 2.2 m_A, \qquad
w   = 0.6 r^M_A - 2.2 m_A, 
\nonumber \\
q   &=& 1, \qquad
s   = 2.8 ,
\ea
is designed such that $Q_A$ is unity in the buffer zone and zero near
black hole A (see Fig.~\ref{transition-func}).
This means that using Eq.~(\ref{InertialKStoCorotIso})
we can transform from Kerr-Schild coordinates (labeled by a hat), to
coordinates (labeled by a tilde) which are corotating isotropic
coordinates in the buffer zone (where $Q_A=1$), but which are equal to
Kerr-Schild coordinates at the black hole horizons (where $Q_A=0$).
The function $Q_A$ is chosen carefully so that the transformed
coordinate system is identically Kerr-Schild at the horizon
($\tilde{r}_A=2m_A$), while it is equal to isotropic coordinates in
the buffer zone where we perform the matching. We can adjust, as
usual, how fast we transition by changing the parameters of $Q_A$, but
we are constrained to having the metric completely in isotropic
coordinates in the buffer zone, in order for the matching to be valid.

The inner zone metric then becomes 
\be
g_{\tilde{\mu}\tilde{\nu}}^{(1)} = g_{\hat{\alpha}\hat{\beta}}^{(1)}
J^{\hat{\alpha}}_{\tilde{\mu}} J^{\hat{\beta}}_{\tilde{\nu}},
\ee
where, in analogy to Eq.~(\ref{jacobian1}), the Jacobian matrix is
given by 
\be
J^{\hat{\alpha}}_{\;\tilde{\mu}} 
= \frac{\partial x^{\hat{\alpha}}}{\partial x^{\tilde{\mu}}}.
\ee
This Jacobian can be computed by taking derivatives of
Eq.~(\ref{InertialKStoCorotIso}).

\begin{figure}
\includegraphics[scale=0.45,clip=true]{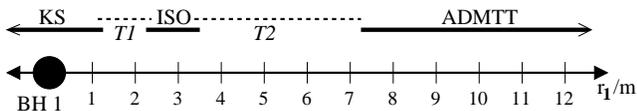}
\caption{\label{transitions} Schematic drawing of the different
metrics, coordinate systems and transition windows. The solid line
represents the distance $r_1$ (in ADMTT coordinates) from BH~1.
The thick solid line labeled by ADMTT delineates the region
where we use the PN near zone solution in ADMTT coordinates.
The other two thick solid lines labeled by ISO and KS, 
show the region in which we use the inner zone solution.
The labels here stand for the isotropic (ISO) and Kerr-Schild (KS)
coordinates we have used in the inner zone before applying the
matching coordinate transformation (\ref{fulltransf-nokappa}). 
The dotted line $T1$ corresponds to the region where the transition 
function $Q_1$ (see Fig.~\ref{transition-func})
of Eqs.~(\ref{Q-trans}) and (\ref{Q-trans-pars})
is between zero and one, so that the coordinate transformation
of Eq.~(\ref{InertialKStoCorotIso}) yields non-trivial results.
I.e. in that region the inner zone metric is given in coordinates that are a
mixture of Kerr-Schild and isotropic coordinates, while to the left and
right of $T1$ it is given in Kerr-Schild and isotropic coordinates,
respectively. Using the definitions (\ref{Q-trans}) and
(\ref{Q-trans-pars}) for the transition function $Q_1$, we find that the
inner and outer radii of $T1$ are $2.2m_A$ and $0.6r^M_A$, where $r^M_A$
given in Eq.~(\ref{Wolfs_rM}).
The dotted line labeled $T2$ indicates the transition window
of the transition function $F_1$ (see Fig.~\ref{transition-func})
 of Eq.~(\ref{transF}) with parameters given
by Eqs.~(\ref{transFpars}) and (\ref{Wolfs_rM}). In $T2$ we use a weighted
average of both the inner zone metric (transformed into ADMTT coordinates
using Eq.~(\ref{fulltransf-nokappa})) and the near zone ADMTT metric.
From Eq.~(\ref{transFpars}) we see that the inner and outer radii
of $T2$ are $0.4r^M_A$ and $3.9r^M_A$.
}
\end{figure}

With this new coordinate transformation, the global metric contains 
several transition functions. A schematic drawing of these transitions is
presented in Fig.~\ref{transitions}.
In this drawing, the black dot represents BH~1 and the solid black
line is the radial direction, where the horizon, for example, is
located at $\tilde{r}_1=2m_1$.  Fig.~\ref{transitions} also shows the
different coordinate systems used, where KS stands for Kerr-Schild,
ISO for isotropic coordinates and ADMTT for the PN near zone. The
dotted line shows the region where the transitions take place: $T1$ is
the transition between KS to ISO and $T2$ is the transition produced by
the matching coordinate transformation. Observe that $T1$ is chosen so
that the $4$-metric is in KS coordinates everywhere near and inside
the horizon, while it is completely in ISO coordinates where $T2$
begins. This restriction makes the transition window of $T1$ narrow. On
the other hand, the transition window of $T2$ is restricted only by the
size of the buffer zone and, thus, is chosen to be wider.

\subsection*{Effect of the new coordinates}

In this subsection, we describe the advantages of the new coordinate
system.  In these coordinates, the new global 4-metric is much better
behaved close to the horizons, where it is of Kerr-Schild form. Thus,
the lapse remains positive definite through the horizon. The lapse for
an equal mass system along the positive x-axis in the new coordinates
is shown in Fig.~\ref{lapse-collapse} (dash-dash-dotted line) together
with the lapse in the old ADMTT coordinates (solid line). The inner
zone lapse in new coordinates is approximately equal to the
Kerr-Schild lapse inside the horizon, while it smoothly approaches the
lapse of isotropic coordinates in the buffer zone and the ADMTT lapse
in the near zone.

\begin{figure}
\includegraphics[scale=0.33,clip=true]{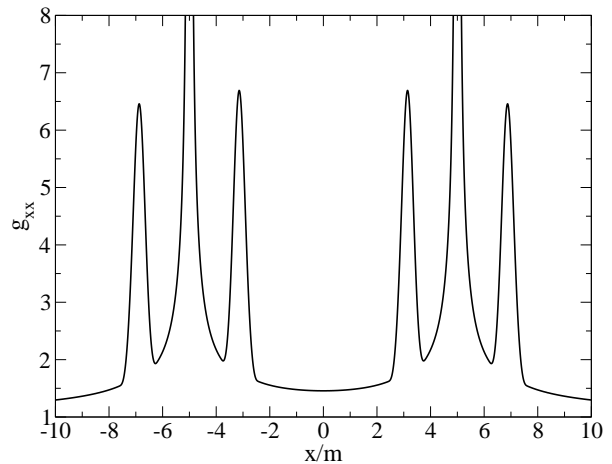}
\caption{\label{gxx-new} 
  This figure shows the $xx$-component of the global metric in new
  coordinates for an equal-mass system along the x-axis.}
\end{figure}

Another quantity that changes in the new coordinates is the spatial
metric. Fig.~\ref{gxx-new} shows the $xx$-component of the new global
metric along the $x$-axis for an equal mass system.
The large humps on either side of each black hole in
Fig.~\ref{gxx-new} are produced by derivatives of the coordinate
transformation (\ref{InertialKStoCorotIso}), which contains the
transition function $Q_A$ that changes rapidly from zero to unity in a
small region of width $w = 0.6 r^M_A - 2.2 m_A$.  Had we chosen a
wider transition window $w$ to transition, this derivative would become
smaller and the hump would be smoothed out.  However, we need the
metric to be completely in isotropic coordinates in the buffer zone
where we perform the matching (approximately at $\bar{r}_A = 6 m_A $).
Therefore, we are constrained to have a narrow transition window $w$,
which then produces large derivatives of the transition functions and
humps in the metric. Note, however, that these humps are
{\textit{not}} spurious gravitational radiation. They simply arise
because of performing a coordinate transformation. Therefore, due to the
inherent diffeomorphism invariance of General Relativity, the physical
content of the data will not be altered by the coordinate
transformation. In addition, if we choose a larger black hole
separation, $r^M_A$ increases and we obtain a wider transition window to
transition, and hence the humps become smaller.

\begin{figure}
\includegraphics[scale=0.33,clip=true]{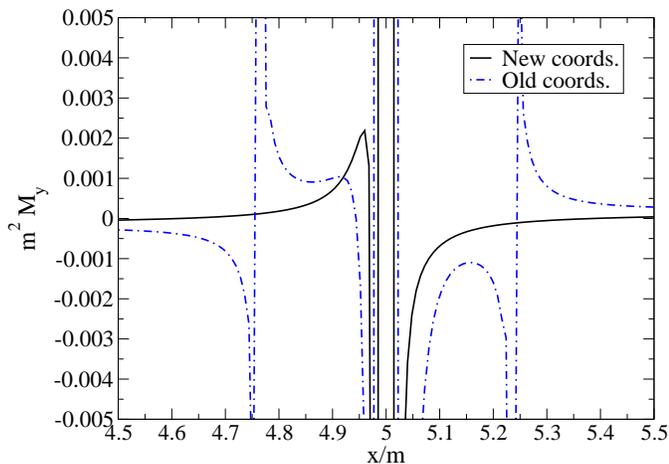}
\caption{\label{py-cons-new-zoom} 
  Plot of the y-component of the momentum constraint violation for an
  equal mass binary near BH~1 in new coordinates (black solid line)
  and in ADMTT coordinates (dot-dashed line).  Observe that in new
  coordinates there is no divergence until the singularity.}
\end{figure}

Since the inner zone lapse in this new coordinates is now a positive
definite function, we expect the extrinsic curvature and the momentum
constraint violation to be small and well-behaved across the horizon.
In Fig.~\ref{py-cons-new-zoom} we compare the $y$-component of the
momentum constraint violation for an equal mass binary in this new
coordinate system to that in the old ADMTT coordinates near BH $1$.
Observe that the constraint violation is finite everywhere, except
near the singularity. With this new coordinate system, excision is now
possible, since the curvature does not blow up until close to the
physical singularity.
We have zoomed to a region $0.5 m$ away from the singularity of BH~1
to distinguish the behavior of the violation better. As we can observe
from the figure, the ADMTT constraint violation is identical to the
violation in the new coordinates away from the horizon. However,
near the horizon there are spikes when we use the old
coordinates. These spikes are poorly resolved in this figure, but we
have checked that they are indeed divergences. Observe that these
spikes are not present in the new coordinates. 

\begin{figure}
\includegraphics[scale=0.33,clip=true]{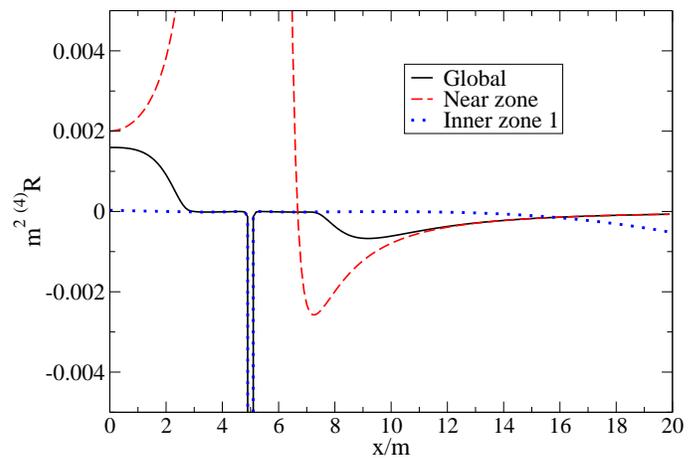}
\caption{\label{Ricciscalar_vs_x} 
This figure shows the Ricci scalar at $t=0$ along the $x$-axis for the 
inner (dotted line), near (dashed line) and global (solid line) 4-metrics.}
\end{figure}
Since our method really yields a 4-metric we can also compute the Ricci
tensor. This tensor should be close to zero for an approximate vacuum
metric. Figure~\ref{Ricciscalar_vs_x} shows the Ricci scalar along the
$x$-axis at $t=0$. One can see that it has the same qualitative features as
the Hamiltonian and momentum constraints. I.e. apart from the singularity
at the center of the black hole, the largest violations occur in the buffer
zone. Notice however, that unlike in the case of metric (see
Fig.~\ref{gxx-new}) the coordinate transformation
(\ref{InertialKStoCorotIso}), with the transition function $Q_A$ does not
produce large humps in the Ricci scalar. This confirms that the large humps
in the metric are a pure coordinate effect, that has no influence on
coordinate invariant quantities such as the Ricci scalar.

\section{Conclusions}
\label{conclusion}

We have constructed approximate initial data for non-spinning black hole 
binaries by asymptotically matching the 4-metrics of two tidally perturbed
Schwarzschild solutions in isotropic coordinates to a
post-Newtonian metric in ADMTT coordinates. 
The two perturbed Schwarzschild metrics are valid only close to each
black hole (in the inner zones) and contain tidal deformations
which are correct up to $O(r_A/b)^2$. The ADMTT post-Newtonian metric
we use is formally correct only up to $O(m/b)^{3/2}$ and valid only
in the near zone, and not close to the black holes.
However, by adding certain higher order terms to the ADMTT metric
we bring it into a form that is close to the Schwarzschild metric
plus some artificial perturbation.
This means that instead of blowing up in an uncontrolled way,
the ADMTT metric approaches the Schwarzschild metric in isotropic
coordinates when we approach one of the black holes. 
Since the two tidally perturbed
Schwarzschild solutions, which are actually valid close to
each black hole are also given in isotropic coordinates, both
metrics agree at leading order even before asymptotic matching is performed.

The procedure we have used to achieve asymptotic matching closely
follows the calculation of paper $1$, and is performed in the so
called buffer zone where both the post-Newtonian and the perturbed
Schwarzschild approximations hold.  The result is that both metrics
agree in the buffer zone, up to the errors in the approximations.
However, since both approximations are similar to Schwarzschild in
isotropic coordinates, matching yields much better results than in
paper $1$, where harmonic coordinates were used for the post-Newtonian
4-metric.  In addition, after matching in the buffer zone the two
metrics are also very similar near the black hole horizons, even
though the post-Newtonian metric is formally not valid there. The
biggest deviation of the post-Newtonian 4-metric from the correct
perturbed Schwarzschild solution near the horizon occurs in the
$g_{0i}$ or shift components at $O(m/b)^{3/2}$.  At this same order
similar deviations are also observable in the extrinsic curvature of
the $t=0$ slice.

The resulting global piece-wise 4-metric is made formally $C^{\infty}$
by the use of smoothing functions. These functions are chosen such
that the smoothed global metric has errors on the order of the error
introduced by the more accurate of the two approximations we match.
This smoothing procedure works much better than in paper $1$, because
the tidally perturbed Schwarzschild solutions and the post-Newtonian
metric are much closer to each other and diverge at similar locations.
The smoothed global metric is obtained in ADMTT coordinates, and thus
very similar to isotropic coordinates near each black hole.  Hence our
coordinates are not horizon penetrating and, for example, result in a
lapse that goes through zero close to the horizons.  Since such
coordinates may be problematic for certain numerical simulations, we
construct an additional coordinate map.  With this map one can
transform the 4-metric from global ADMTT coordinates, obtained through
asymptotic matching, to new coordinates which are similar to
Kerr-Schild coordinates near each black hole, but which remain ADMTT
further away from the black holes.  These new coordinates are horizon
penetrating and lead to a lapse which is everywhere positive on the
$t'=0$ slice. The same map can in principle also be applied to the
data computed in paper $1$.

For both, global ADMTT and new coordinates we have constructed
initial data sets on the respective $t'=0$ slices.
These initial data are then used to compute Hamiltonian and momentum
constraint violations. We find that the
initial data found in this paper is closer to the
constraint hypersurface that the data presented in paper $1$, 
making it perhaps more amenable to numerical simulations.

In conclusion, we have used the method of asymptotic matching to construct
improved approximate initial data for non-spinning black hole binaries.
Future work will concentrate on applying this method to spinning binary
systems. This problem is of fundamental importance to astrophysics and
gravitational wave detection, since most black holes are believed to be
rotating. Using the results of~\cite{Yunes:2005ve}
the matching of a tidally perturbed Kerr black hole
to a post-Newtonian metric should be possible.
Other work will concentrate on repeating this analysis to higher
post-Newtonian order to explicitly incorporate the effect of
gravitational waves.

\begin{acknowledgments}

We would like to thank the University of Jena for their hospitality. 
We would also like to thank Ben Owen and Bernd Br\"ugmann for useful 
discussion and comments.  
Nicolas Yunes acknowledges the support of the Institute for
Gravitational Physics and Geometry and the Center for Gravitational
Wave Physics, funded by the National Science Foundation under
Cooperative Agreement PHY-01-14375. This work was also supported by
NSF grants PHY-02-18750, PHY-02-44788, and PHY-02-45649.
Wolfgang Tichy acknowledges partial support by the
National Computational Science Alliance under 
Grants PHY050012T, PHY050015N, and PHY050016N.

\end{acknowledgments}

\appendix

\section{Jacobian}
\label{appendix_Jac}
In this section we present explicit formulas for the Jacobian of the
transformations given in Eq.~(\ref{fulltransf-nokappa}). These formulas
are long and complicated, but, if the reader is interested in
implementing them, we can provide a Maple file with explicit
expressions for them. The Jacobian is given by 

\begin{widetext}
\ba
J^{\bar{t}}_{t} &=& 1+ \frac{m_2}{b} \, \left( -1+{\frac {{ x_1}}{b}}
\right),
\nonumber \\
J^{\bar{t}}_{x} &=& {\frac {t{ m_2}}{{b}^{2}}},
\nonumber \\
J^{\bar{x}}_{t} &=& { x_1}\, \left( {\frac {{ m_2}\,y}{{ x_1}\,{b}^{2}}}- \left( {
\frac {{ m_2}}{b}} \right) ^{3/2}\sqrt {{\frac {m}{{ m_2}}}}t
 \left\{ \frac{x_1}{b^2} \, \left[ \frac{1}{2}\, \frac{m_1}{m}\, \left( {\frac {{b}^{3}}{
 r_1^{3}}}-1+3\,{\frac {{m_1}}{m}} \right) - 4 \right] 
-\frac{3}{2}\,{\frac {{y}^{2}{x_1}\,{ m_1}\,b}{ r_1^{5}m}}
\right.\right.
\nonumber \\ 
&& \left. \left.
+ \left( 1-{\frac {{ m_2}}{m}}-\frac{1}{2}\,{\frac {{ m_1}}{m}}
 \right) {b}^{-1} \right\} {{ x_1}}^{-1} \right), 
\nonumber \\
J^{\bar{x}}_{x} &=& 1+ \frac{m_2}{b}\, \left( 1-\frac{1}{2}\,{\frac {{
        x_1}}{b}}+\frac{1}{2}\,{\frac {{y}^{2}+ 
{z}^{2}+2\,{t}^{2}}{{ x_1}\,b}} \right) - \left( {\frac {{
 m_2}}{b}} \right) ^{3/2}\sqrt {{\frac {m}{{ m_2}}}}t \left\{ \frac{y
        x_1}{b^2} \left[ \frac{1}{2}\,{ m_1}\, \left( {\frac {{b}^{3}}{ 
       r_1^{3}}}-1 
\right. \right. \right.
\nonumber \\ \nonumber 
&& \left. \left. \left. 
+3\,{\frac {{ m_1}}{m}} \right) {m}^{-1}-4 \right] + \frac{y}{b} \left(
        1-{\frac {{ m_2}}{m}}-\frac{1}{2}\,{\frac {{ m_1}}{m}} \right)
        \right\} {{ x_1}}^{-1}+ { x_1}\, \left( \frac{m_2}{b}\,
        \left( -\frac{1}{2 b} - \frac{1}{2}\,{\frac {{y}^{2
        }+{z}^{2}+2\,{t}^{2}}{{{ x_1}}^{2}b}} \right)  
\right.
\nonumber \\
&& \left. 
- \left( {\frac {{ m_2}}{b}} \right) ^{3/2}\sqrt {{\frac {m}{{ m_2}}}}t
 \left\{ \frac{y}{b^2} \left[ \frac{1}{2}\,\frac{m_1}{m} \left( {\frac {{b}^{3}}{
           r_1^{3}}}-1+3\,{\frac {{ m_1}}{m 
}} \right) - 4 \right]  -\frac{3}{2}\,{\frac {y{{ x_1}}^{2}{m_1}\,b}{
        r_1^{5}m}}  \right\} {{ x_1}}^{-1}
\right. 
\nonumber \\
&& \left. 
 + \left( {\frac {{ m_2}}{b}} \right) ^{3/2}
\sqrt {{\frac {m}{{ m_2}}}}t \left\{ \frac{y x_1}{b^2} \, \left[ \frac{1}{2}\,{ m_1}
\, \left( {\frac {{b}^{3}}{ r_1^{3}}}-1+3\,{\frac {{ m_1}}{m}} \right) {m}^{-1}
-4  \right] + \frac{y}{b} \left( 1-{\frac {{ m_2}}{m}}-\frac{1}{2}\,{\frac {{ 
m_1}}{m}} \right) \right\} {{ x_1}}^{-2} \right), 
\nonumber \\
J^{\bar{x}}_{y} &=& { x_1}\, \left( {\frac {{ m_2}\,y}{{ x_1}\,{b}^{2}}}- \left( {
\frac {{ m_2}}{b}} \right) ^{3/2}\sqrt {{\frac {m}{{ m_2}}}}t
 \left\{ \frac{x_1}{b^2} \left[ \frac{1}{2}\,\frac{m_1}{m}\, \left(
       {\frac {{b}^{3}}{r_1^{3}}}-1+3\,{\frac {{m_1}}{m}} \right) - 4
   \right] -\frac{3}{2}\,{\frac {{y}^{2}{ x_1}\,{ m_1}\,b}{ r_1^{5}m}}
\right. \right.
\nonumber \\
&& \left. \left. 
+ \left( 1-{\frac {{ m_2}}{m}}-\frac{1}{2}\,{\frac {{ m_1}}{m}}
 \right) {b}^{-1} \right\} {{ x_1}}^{-1} \right),
\nonumber \\
J^{\bar{x}}_{z} &=& { x_1}\, \left[ {\frac {{ m_2}\,z}{{ x_1}\,{b}^{2}}}+\frac{3}{2}\,
 \left( {\frac {{ m_2}}{b}} \right) ^{3/2}\sqrt{\frac{m}{m_2}} \frac{t
 y m_1 b z}{r_1^5 m} \right], 
\nonumber \\
J^{\bar{y}}_{t} &=& y \left( \sqrt {{\frac {{ m_2}}{b}}}\sqrt {{\frac {{ m_2}}{m}}}{y}
^{-1}- \left( {\frac {{ m_2}}{b}} \right) ^{3/2}\sqrt {{\frac {m}{{
 m_2}}}} \left\{ \frac{{z}^{2}}{b^2} \left( \frac{7}{4}\,{\frac {{ m_1}}{m}}-{\frac {{
 m_2}}{m}}-\frac{3}{2} \right) + \frac{{y}^{2}}{b^2} \left[ \frac{1}{2}\,\frac{m_1}{m}\,
 \left( \frac{5}{2} +{\frac {{b}^{3}}{ r_1^{3}}} \right) 
-\frac{1}{2}-{\frac {{ m_2}}{m}} \right] 
\right. \right. 
\nonumber \\
&& \left. \left. 
+ \frac{{{x_1}}^{2}}{b^2} \left( 2\,{\frac {{ m_2}}{m}}-\frac{7}{2}\,{\frac {{
 m_1}}{m}}+\frac{7}{2} \right) + \frac{x_1}{b}\, \left( -1+{\frac {{ m_2
}}{m}} +\frac{7}{2}\,{\frac {{ m_1}}{m}} \right) + \frac{1}{3}\,{\frac {{t}^{2
}}{{b}^{2}}}
+\frac{7}{2}\,\frac{m_1}{m}\, \left( -1+{\frac {b}{r_1}} \right) +
\frac{3}{2}-\frac{1}{2}\,{\frac {{ m_1}\,{  
m_2}}{{m}^{2}}}
\right. \right.
\nonumber \\
&& \left. \left.
-{\frac {{ m_2}}{m}} \right\} {y}^{-1} - \frac{2}{3} \, \left( {
\frac {{ m_2}}{b}} \right) ^{3/2} \frac{{t}^{2}}{b^2 y}\sqrt {{\frac {m}{{ m_2}}}
} \right), 
\nonumber \\
J^{\bar{y}}_{x} &=& y \left\{ -{\frac {{ m_2}}{{b}^{2}}}- \left( {\frac {{ m_2}}{b}}
 \right) ^{3/2}\sqrt {{\frac {m}{{ m_2}}}}t \left[ -\frac{3}{2}\,{\frac {{y}
^{2}{ x_1}\,{ m_1}\,b}{ r_1^{5}m}}+ 2\, \frac{x_1}{b^2} \, \left(
 2\,{\frac {{ m_2}}{m}}-\frac{7}{2}\, 
{\frac {{ m_1}}{m}}
+\frac{7}{2} \right) + \left( -1+{\frac {{ m_2}
}{m}}+\frac{7}{2}\,{\frac {{ m_1}}{m}} \right)
 {b}^{-1}
\right. \right.
\nonumber \\
&& \left. \left. 
-\frac{7}{2}\,{\frac {{ m_1 }\,b{ x_1}}{ r_1^{3}m}}  \right] {y}^{-1} \right\}, 
\nonumber \\
J^{\bar{y}}_{y} &=& 1+\sqrt {{\frac {{ m_2}}{b}}}\sqrt {{\frac {{
        m_2}}{m}}} \frac{t}{y} + \frac{m_2}{b} \left( 1-{\frac {{
        x_1}}{b}} \right) - \left( {
\frac {{ m_2}}{b}} \right) ^{3/2}\sqrt {{\frac {m}{{ m_2}}}} \frac{t}{y}
 \left\{ \frac{{z}^{2}}{b^2} \left( \frac{7}{4}\,{\frac {{ m_1}}{m}}-{\frac {{ m_2}}{m}
}-\frac{3}{2} \right) + \frac{{y}^{2}}{b^2} \left[ \frac{1}{2}\, \frac{m_1}{m}\, 
\right. \right.
\nonumber \\ 
&& \left. \left.
\left( \frac{5}{2} +{\frac {{b}^{3}}{ r_1^{3}}}\right)
        -\frac{1}{2}-{\frac {{ m_2}}{m}} \right] + 
        \frac{{{x_1}}^{2}}{b^2} \left( 2\,{\frac {{
        m_2}}{m}}-\frac{7}{2}\,{\frac {{ m_1}}{m}}+\frac{7}{2} 
 \right) + \frac{x_1}{b}\, \left( -1+{\frac {{ m_2}}{m}}
+\frac{7}{2}\,{\frac {{ m_1}}{m}} \right) +\frac{1}{3}\,{\frac
        {{t}^{2}}{{b}^{2}}} 
\right. 
\nonumber \\
&& \left.
+ \frac{7}{2}\, \frac{m_1}{m} \left( -1+{\frac {b}{r_1}} \right)
        +\frac{3}{2}-\frac{1}{2}\,{\frac {{ m_1}\,{
        m_2}}{{m}^{2}} }-{\frac {{ m_2}}{m}} \right\} 
+y \left( -\sqrt {{\frac {{m_2}}{b}}}\sqrt {{\frac {{
        m_2}}{m}}} \frac{t}{{y}^{2}} - \left( {\frac {{
 m_2}}{b}} \right) ^{3/2}\sqrt {{\frac {m}{{ m_2}}}} \frac{t}{y} 
\right.
\nonumber \\
&& \left. 
\left\{ 2\, \frac{y}{b^2}  \left[ \frac{1}{2}\,\frac{m_1}{m}\, \left( \frac{5}{2}+{\frac
        {{b}^{3}}{r_1^{3}}} \right) - \frac{1}{2}-{\frac {{  m_2}}{m}}
        \right] -\frac{3}{2}\,{\frac {{y}^{3}{ m_1}\,b}{ r_1^{5}m}}
-\frac{7}{2}\,{\frac {{ m_1}\,by}{ r_1^{3}m}}  \right\} +
        \left( {\frac {{ m_2}}{b}} \right) ^{3/2}\sqrt {{ 
\frac {m}{{ m_2}}}}t \left\{ \frac{{z}^{2}}{b^2} \left( \frac{7}{4}\,{\frac {{ m_1}}{m}
}-{\frac {{ m_2}}{m}}
\right. \right. \right.
\nonumber \\
&& \left. \left. \left. 
 -\frac{3}{2} \right)  + \frac{{y}^{2}}{b^2} \left[ \frac{1}{2}\,{ 
m_1}\, \left( \frac{5}{2}
+{\frac {{b}^{3}}{ r_1^{3}}} \right) {m}^{-1}-\frac{1}{2}-{\frac {{ m_2}}{m}}
 \right] + \frac{{{x_1}}^{2}}{b^2} \left( 2\,{\frac {{ m_2}}{m}}-\frac{7}{2}\,
{\frac {{ m_1}}{m}}
\right. \right. \right.
\nonumber \\
&& \left.  \left. \left. 
+\frac{7}{2} \right) + \frac{x_1}{b}\, \left( -1+{
\frac {{ m_2}}{m}}+\frac{7}{2}\,{\frac {{ m_1}}{m}} \right) 
+\frac{1}{3}\,
{\frac {{t}^{2}}{{b}^{2}}}+\frac{7}{2}\,\frac{m_1}{m}\, \left( -1+{\frac {b}{
        r_1}} \right) + \frac{3}{2} 
-\frac{1}{2}\,{\frac {{ m_1}\,{ m_2}}{{m}^{2}}} -{\frac {{ m_2}}{m}} \right\}
{y}^{-2} 
 \right), 
\nonumber
\ea
\ba
J^{\bar{y}}_{z} &=& - \left( {\frac {{ m_2}}{b}} \right) ^{3/2}t\sqrt {{\frac {m}{{ 
m_2}}}} \left[ 2\, \frac{z}{b^2} \left( \frac{7}{4}\,{\frac {{ m_1}}{m}}-{\frac {{ m_2}}
{m}}-\frac{3}{2} \right) - \frac{3}{2}\,{\frac {{y}^{2}{ m_1}\,bz}{ 
  r_1^{5}m}}-\frac{7}{2}\,{\frac {{ m_1}\,
bz}{ r_1^{3}m}} \right],
\nonumber \\
J^{\bar{z}}_{t} &=& -z \left( {\frac {{ m_2}}{b}} \right) ^{3/2}
\frac{y}{b^2} \sqrt {{\frac {m}{{ m_2}}}} \left[ \frac{1}{2}\, \frac{m_1}{m}\,
  \left( {\frac {{b}^{3}}{ r_1^{3}}}-1 \right) + 1 \right],
\nonumber \\
J^{\bar{z}}_{x} &=& z \left[ -{\frac {{ m_2}}{{b}^{2}}}+\frac{3}{2}\,
  \left( {\frac {{ m_2}}{b }} \right) ^{3/2}yt\sqrt {{\frac {m}{{
          m_2}}}} \frac{{ m_1}\,b{ x_1}}{r_1^{5}{m}^{1}} \right], 
\nonumber \\
J^{\bar{z}}_{y} &=& z \left\{ - \left( {\frac {{ m_2}}{b}} \right)
  ^{3/2} \frac{t}{b^2} \sqrt {{\frac {m}{{ m_2}}}} \left[
    \frac{1}{2}\, \frac{m_1}{m}\, \left( {\frac {{b}^{3}}{ 
  r_1^{3}}}-1 \right) + 1 \right] +\frac{3}{2}\, \left( {\frac
{{ m_2}}{b}} \right) ^{3/2}{y} ^{2}t\sqrt {{\frac {m}{{ m_2}}}}
\frac{{m_1}b}{r_1^{5}{m}} \right\}, 
\nonumber \\
J^{\bar{z}}_{z} &=& 1+ \frac{m_2}{b}\, \left( 1-{\frac {{ x_1}}{b}}
\right) - \left( { \frac {{ m_2}}{b}} \right) ^{3/2}\frac{yt}{b^2} \sqrt {{\frac
    {m}{{ m_2}}}}  \left[ \frac{1}{2}\, \frac{m_1}{m}\, \left( {\frac
      {{b}^{3}}{ r_1^{3}}}-1 \right) + 1 \right]
+\frac{3}{2}\,{z}^{2} \left( {\frac {{ m_2}}{b}} \right) ^{3/2}
yt\sqrt {{\frac {m}{{ m_2}}}} \frac{{m_1}\,b}{r_1^{5}{m}}. 
\ea
\end{widetext}
where terms not listed here are zero. 

This is the Jacobians of the transformations that allows us to
construct a global metric by transforming the inner zone metrics in
the buffer zone. Note that this is the Jacobian for the transformation
that is valid in buffer zone $1$ (${\cal{O}}_{13}$.) In order to
obtain the Jacobian for the transformation in the other buffer zone
(${\cal{O}}_{23}$), one can apply the substitutions of
Eq.~(\ref{symmetry}).

\bibliography{paper.bib}

\end{document}